\documentclass[12pt]{amsart}
\usepackage{a4wide}
\usepackage{amssymb}
\usepackage{url}

\newtheorem{theorem}{Theorem}[section]

\newtheorem{lemma}[theorem]{Lemma}

\theoremstyle{definition}
\newtheorem{definition}[theorem]{Definition}
\newtheorem{remark}[theorem]{Remark}

\numberwithin{equation}{section}

\begin{document}

\title[Portfolio choice under monotone preferences - stochastic factor case]{Continuous-Time Portfolio Choice Under Monotone Mean-Variance Preferences{\textemdash}Stochastic Factor Case}

\author{Jakub Trybu{\l}a}

\address{\noindent Jakub Trybu{\l}a, \newline \indent Department of Mathematics \newline \indent Uniwersity of Economics \newline \indent   Rakowicka 27, \newline \indent  31--510  Krak{\'o}w, Poland}

\email{jakubtrybula@gmail.com}

\author{Dariusz Zawisza}
\address{\noindent Dariusz Zawisza, \newline \indent Institute of Mathematics \newline \indent Faculty of Mathematics and Computer Science \newline \indent  Jagiellonian University \newline \indent{\L}ojasiewicza  6 \newline \indent 30-348 Krak{\'o}w, Poland }

\email{dariusz.zawisza@im.uj.edu.pl}

\subjclass[2010]{91G10; 91A15; 91A23; 93E20}

\keywords{continuous optimization; stochastic control; stochastic factor model, Heston model}

\date{\today}
\maketitle

\begin{center}
First arXiv version: 13 March 2014 \\
This version: Accepted manuscript\footnote{History:Received--December 22, 2015; Accepted--May 20, 2018;
Published Online--May 29, 2019 } \\
Final version: published in Math. Oper. Res. 44 (2019), 966 -- 987\\
\url{https://doi.org/10.1287/moor.2018.0952} \\
\end{center}

\begin{abstract}
We consider an incomplete market with a non-tradable stochastic factor and a continuous time  investment problem with an optimality criterion based on  monotone mean-variance preferences.  We formulate it as a stochastic differential game problem and use Hamilton-Jacobi-Bellman-Isaacs equations to find an optimal investment strategy and the value function. What is more, we show that our solution is also optimal for the classical Markowitz problem and every optimal solution for the classical Markowitz problem is optimal also for the monotone mean-variance preferences. These results are interesting because the original Markowitz functional is not monotone, and it was observed that in the case of a static one-period optimization problem the solutions for those two functionals are different. In addition, we determine explicit Markowitz strategies in the square root factor models.
\end{abstract}

\section{Introduction.}\label{intro}

Since Markowitz published his famous paper \cite{Markowitz}, the mean-variance criterion has became a very popular topic in the investment literature. First, the problem has been solved for a myopic investor in the static optimization framework. Then, it has been extended and has been solved in the multi-period framework, when intertemporal trading is allowed. The exact solution can be found in Li and Ng \cite{Li} (discrete time setting) and Zhou and Li \cite{Zhou} (continuous time framework).  

On the other hand, it is commonly accepted that a proper decision functional should reflect the fact that the main motivation of a rational  investor is to earn money and thus, when choosing between two prospects (investment returns) $X$ and $Y$, such that $X \leq Y$, the investor will always choose $Y$. This type of behavior is often formulated in the form of a monotonicity condition. Namely, we say that the functional $\rho$ satisfies a monotonicity condition if the relation $X\leq Y $ implies $\rho(X) \leq \rho(Y)$. Note that in case of portfolio optimization problem, the use of a not monotone functional may lead to irrational decisions  because there might be a better (greater) prospect which is excluded by the functional. Therefore, such axiom is usually incorporated in the majority of modern theories of a rational investor behavior (e.g. the expected utility theory - von Neumann and Morgenstern \cite{Neumann}, the dual theory of choice - Yaari \cite{Yaari}, the max--min theory - Gilboa and Schmeidler \cite{Gilboa}, dynamic variational preferences - Maccherroni et al.  \cite{Maccheroni}, coherent and convex risk measures - Artzner et al. \cite{Artzner}, F\"ollmer and  Schied \cite{Follmer}). Nevertheless, it is well known that the mean-variance functional is not monotone.  For this reason Maccheroni et al. \cite{Maccheroni} created a new class of  preferences that {\it coincide with mean-variance preferences on their domain of monotonicity but differ where the mean-variance preferences fail to be monotone:}
\begin{equation} \label{functional}
V_{\theta}(X):=\inf _{Q \in \mathcal{Q}}\left\{\mathbb{E}^{Q}\left[X\right]+\frac{1}{2\theta}C(Q|P)\right\},\qquad X\in\mathcal{L}^{2}(P),
\end{equation}
where $\theta>0$ is a risk aversion coefficient, $P$ is a given probability measure, $\mathcal{L}^{2}(P)$ is the set of square-integrable random variables with respect  to the measure $P$, $\mathcal{Q}$ is the class of all probability measures, such that
\[
C(Q|P):=\left\{ 
\begin{aligned}
&\mathbb{E}^{P}\left[\left(\frac{dQ}{dP}\right)^{2}\right]-1,\qquad&\text{if}\quad Q\ll P,\\
&+\infty,\qquad&\text{otherwise}.
\end{aligned}
\right.
\]
Moreover, they have shown that {\it the functional associated with this class of preferences is the best approximation of the mean-variance functional among those which are monotonic}. For more details about monotone mean-variance preferences and their other advantages over mean-variance preferences, we refer to Maccheroni et al. \cite{Maccheroni}.

It is also worth mentioning that (\eqref{functional}) can be rewritten as
\begin{equation} \label{shown}
V_{\theta}(X)=-\Lambda_{\theta}(X)-\frac{1}{2\theta},
\end{equation}
where
\[
\Lambda_{\theta}(X):= \sup _{Q \in \mathcal{Q}}\mathbb{E}^{Q}\left[-X-\frac{1}{2\theta}\frac{dQ}{dP}\right], \qquad X\in\mathcal{L}^{2}(P),
\]
so additional motivation to investigate further properties of such type performance criterion lies in the fact that the above functional satisfies the following axioms:
\begin{itemize} 
 \item[]\emph{Convexity:} If $\alpha \in (0,1)$, then $\Lambda_{\theta}(\alpha X+ (1-\alpha)Y) \leq \alpha \Lambda_{\theta}(X) + (1-\alpha) \Lambda_{\theta}(Y)$.
 \item[]\emph{Monotonicity:} If $X\leq Y $,  then $\Lambda_{\theta}(Y) \leq \Lambda_{\theta}(X)$.
 \item[]\emph{Translation invariance:} If $\beta\in\mathbb{R}$, then $\Lambda_{\theta}(X+ \beta)=\Lambda_{\theta}(X)-\beta$.
\end{itemize} 
Namely, $\Lambda_{\theta}(X)$ is a convex risk measure (see  F\"{o}llmer  and Schied \cite{Follmer} or Frittelli and Rosazza-Gianin \cite{Frittelli}). Note that this type of measures is commonly accepted as a tool for assigning the risk to  financial positions, so the minimization  of $\Lambda_{\theta}(X)$ can be interpreted as  searching for the risk minimizing portfolio. In this case the function $C(Q|P)$ is called a penalty function. 

The portfolio optimization problem based on the functional $V_{\theta}(X)$ has been solved by Macheroni et al. \cite{Maccheroni} in the static setting. Since this functional has never been investigated in intertemporal setting, it is important to describe an optimal financial strategy which the investor can follow in order to maximize $V_{\theta}(X)$ and  investigate whether the same differences between monotone and not monotone counterpart arise in the continuous   
time optimization framework. 

In this paper we assume that an investor has access to the market, where he can freely buy and sell a risk-less bond and a risky asset whose price is a diffusion with dynamics affected by a correlated non-tradable (but observable) stochastic factor. 
Nowadays, stochastic factor models have become very popular in  continuous time portfolio optimization theories. Such models can incorporate many empirical findings about stochastic  market returns, for example stochastic nature of the volatility or factor dependence on the excess return. Many researchers have tried to investigate what is the impact of the factor on the risky asset for various optimization criterions, usually under the hyperbolic/constant absolute risk aversion utility framework. The topic was explored, among others, by  Kim and Omberg \cite{Kim}, Campbell and
Viceira \cite{Campbell}, Fleming and Hern\'{a}ndez \cite{Fleming2}, Liu \cite{Liu},  Taksar and Zeng \cite{Taksar} and  Zariphopoulou \cite{Zari2}. Besides,  stochastic factor models are fundamental examples of incomplete market models and it is important to check various properties of the Markowitz portfolio in the case of these models because they may lead to various paradoxes. For example, as documented in B\"{a}uerle and Grether \cite{Bauerle} and Cui et al. \cite{Cui}, they might be responsible for generating so called free cash flow streams.

The problem of looking for a risk minimizing portfolio,  with various modifications of performance criterion (\eqref{functional}), was considered by many authors. For example Mataramvura and {\O}ksendal \cite{Mataramvura} studied this issue in a jump diffusion setting with a general penalty function of the form
\[
C(Q|P)=h\left( \frac{dQ}{dP} \right).
\]  
The same kind of problem was examined by Elliott and Siu \cite{Elliott4} in the context of an optimal reinsurance problem and by Elliott and Siu \cite{Elliott3, Elliott} in a regime switching market. 

Risk based portfolio problems are also useful for derivative pricing in incomplete markets. Namely, one  possibility is to determine the value by considering the so called  risk indifference price. For more information about an indifference price in a jump diffusion market see {\O}ksendal and Sulem \cite{Oksendal2}, while for the stochastic factor model it is worth to read Elliott and Siu \cite{Elliott5}. It merits mentioning that in many papers the utility function is used to take into account the non-linear form of the investor's satisfaction in  functional (\eqref{functional}). This approach is called the robust utility portfolio optimization and was taken up in Hern\'{a}ndez and Schied \cite{Schied} (stochastic factor model),  {\O}ksendal and Sulem \cite{Oksendal3} (jump-diffusion risk),  Bordigioni et al. \cite{Bordigioni} (more general semimartingale setting).

All of the aforementioned papers examine the problem without presenting any detailed solution for a specific choice of $C(Q|P)$ or consider one specific example using the entropic penalty function of the form
\[
C(Q|P)=\frac{dQ}{dP} \ln\left(\frac{dQ}{dP}\right).
\]
Due to the best of our knowledge the quadratic penalty
\[
C(Q|P)= \mathbb{E}^{P}\left[\left(\frac{dQ}{dP}\right)^{2}\right]-1
\]
has never been studied in detail in the dynamic optimization framework. 

The problem of maximizing (\eqref{functional}) is a max-min problem, hence it naturally forms a stochastic differential game. In the literature there are two main approaches for determining the solution to such games. First of them exploits  maximum principle and Backward Stochastic Differential Equations (BSDE), while the second one is based on the dynamic programming principle  and Hamilton-Jacobi-Bellman equations (Hamilton-Jacobi-Bellman-Isaacs for differential games). 

The latter is more suitable for our approach. In our case the associated HJBI equation can be simplified to a linear form by applying certain transformations.  To complete the reasoning it is sufficient to use a suitable version of a verification theorem. By this method we obtain a formula for the optimal strategy, which is characterized by the aforementioned linear equation. Further, the solution is proved to be optimal in the classical Markowitz framework, which is in opposition to the one-period setting. It seems that the continuous time trading (even in an incomplete market setting) ensures enough flexibility to react faster against changing  conditions  and allows the mean-variance investor to "behaves like a monotone investor".

It should be noticed that our paper is not the first one where such a qualitative difference between discrete and continuous time models is presented. Cui et al. \cite{Cui} have shown that a complete financial market is time-consistent in efficiency in the continuous time framework. It is also worth mentioning that the time inconsistency (see for instance Basak and Chabakauri \cite{Basak}) and the lack of monotonicity are two major drawbacks of the Markowitz optimization, commonly discussed in the literature. Moreover, Kallsen et al. \cite{Kallsen} reported that the expression for the hedging error in the variance optimal hedging problem is much more involved in the discrete than in the continuous time setting. 

The paper is organized as follows. In Section 2 we describe the set up of the problem. We formulate the verification theorem and derive the HJBI equation. In Section 3 we transform our equation to a linear form and prove some useful properties of its solution. In Section 4 we solve an auxiliary portfolio optimization problem and then in Section 5 we compare the result with the solution to the classical mean-variance optimization problem. Finally, in Section 6 we formulate our main theorem. In the last section we use methods from previous chapters to determine explicit formulas for the mean-variance optimal  portfolio in specific stochastic volatility models and we compare it further with the Black-Scholes model.

It is worth to note here that in this work in many places we write about PDEs and talk about solutions.  In this article by a solution to PDE we always mean a classical smooth solution.

\section{General model description.}\label{sec:1}

Let $(\Omega, \mathcal{F}, P)$ be a probability space with a filtration $(\mathcal{F}_{t}, 0 \leq t \leq T)$ possibly enlarged to satisfy the usual assumptions and generated by two independent Brownian motions $(W_{t}^{1}, 0 \leq t \leq T )$, $(W_{t}^{2}, 0 \leq t \leq T )$ defined on $(\Omega, \mathcal{F}, P)$. Suppose that an investor has access to the market, where he can freely buy and sell two securities: a bond $(B_{t},s \leq t \leq T)$ and a share $(S_{t},s \leq t \leq T)$. We assume that the dynamics of $(S_{t},s \leq t \leq T)$  depend on one non-tradable (but observable) external factor  $(Z_{t},s \leq t \leq
T)$. This factor can be used to model stochastic volatility or other varying economic conditions. Processes mentioned above are described by the system
\begin{equation}\label{model}
\left\{
\begin{aligned}
dB_{t} &=r B_{t} dt,  \\
dS_{t} &=\mu(Z_{t}) S_{t}  dt + \sigma(Z_{t}) S_{t}  dW_{t}^{1},   \\
dZ_{t} &=a(Z_{t}) dt + b(Z_{t})(\rho dW_{t}^{1} + \bar{\rho} dW_{t}^{2}),\qquad Z_{s}=z,
\end{aligned}
\right.
\end{equation}
where the coefficients $\sigma>0$, $\mu$, $a$, $b$  have all the required regularity conditions, in order to ensure that a unique strong solution to (\eqref{model}) exists. The interest rate $r>0$ is constant, $\rho \in [-1,1]$ is a correlation coefficient and $\bar{\rho}:=\sqrt{1-\rho^{2}}$.
The coefficients are time-independent only for notational convenience.

\subsection*{Performance functional.}\label{sec:2} In this paper we use the monotone mean-variance preferences of the form (\eqref{functional}). Nevertheless, due to technical difficulties, our problem is solved first for the auxiliary functional given by
\begin{equation}\label{functional2}
 \bar{\Lambda}_{\theta}(X):=\sup _{Q \in  \bar{\mathcal{Q}}}\mathbb{E}^{Q}\left[-X-\frac{1}{2\theta}\frac{dQ}{dP}\right], \qquad X\in\mathcal{L}^{2}(P),
\end{equation}
where:
\begin{itemize}
\item  $\bar{\mathcal{Q}}$ is of the form\footnote{For more information see Hern\'{a}ndez and Schied \cite{Schied}.}
\begin{equation}\label{Q}
 \bar{\mathcal{Q}}:=\left\{Q\sim P \colon \frac{dQ}{dP} =\mathcal{E} \left( \int \eta_{t,1}dW_{t}^{1} + \eta_{t,2}dW_{t}^{2}
\right)_{T}\; , \quad (\eta_{1},\eta_{2}) \in \mathcal{M}\right\};
\end{equation}
\item $\mathcal{E}(\cdot)_{t}$ denotes the Doleans-Dade exponential;
\item $\mathcal{M}$ is the set of all progressively measurable processes $\eta =(\eta_{1},\eta_{2})$ taking values in $\mathbb{R}^{2}$, such that
\[
\mathbb{E}^{P}\left[\left(\frac{dQ^{\eta}}{dP}\right)^{2}\right] < +\infty \quad \text{ and } \quad \mathbb{E}^{P}\left[\frac{dQ^{\eta}}{dP}\right] =1;
\]
\item $Q^{\eta}$ denotes the probability measure determined by $\eta\in \mathcal{M}$. 
\end{itemize} 
Note that the above set of assumptions allows us to change the probability measure by using the Girsanov Theorem and guarantees that our modification of the Maccheroni type objective function is well defined. 

In order to simplify formula (\eqref{functional2}), we define the additional family of stochastic processes $(Y_{t}^{\eta}, s\leq t\leq T)$ which are given by the stochastic differential equations
\[
dY_{t}^{\eta}=\eta_{t,1}Y_{t}^{\eta}dW_{t}^{1} + \eta_{t,2}Y_{t}^{\eta}dW_{t}^{2},\qquad Y_{s}^{\eta}=y>0, \qquad \eta \in \mathcal{M}.
\]
Then, it is easy to see that
\[
Y_{T}^{\eta}=y\frac{dQ^{\eta}}{dP},\qquad \eta\in\mathcal{M} 
\]
and
\[
 \bar{\Lambda}_{\theta}(X)=\sup _{\eta \in \mathcal{M}}\mathbb{E}^{\eta}\left[-X-Y_{T}^{\eta}\right],\qquad X\in\mathcal{L}^{2}(P),
\]
where $\mathbb{E}^{\eta}$ denotes the expectation with respect to the measure $Q^{\eta}$ and
\[
y=\frac{1}{2\theta}.
\]

\subsection*{Formulation of the problem.}\label{sec:3} 

We assume that at any time the investor can decide on the absolute value invested in the risky asset, and the value invested in the bank account. Let  $(\bar{X}^{\bar{\pi}}_{t}, s \leq t \leq T )$ be the investor's wealth process with the following dynamics
\[
d\bar{X}_{t}^{\bar{\pi}}=(r\bar{X}_{t}^{\bar{\pi}}+\bar{\pi}_{t}(\mu(Z_{t})-r))dt+\bar{\pi}_{t}\sigma(Z_{t})dW_{t}^{1},\qquad \bar{X}_{s}^{\bar{\pi}}=\bar{x}>0,
 \]
where $\bar{x}$ denotes a current wealth of the investor, whereas the control $\bar{\pi}_{t}$ can be interpreted as the absolute value invested in $S_{t}$. Note that $\bar{\pi}_{t}$ as well as the portfolio wealth $\bar{X}_{T}^{\bar{\pi}}$ are allowed to be negative. In this work it is convenient to use forward values of $\bar{\pi}_{t}$ and $\bar{X}^{\bar{\pi}}_{t}$. Namely, let
\[
\pi_{t}:=e^{r(T-t)}\bar{\pi}_{t}\qquad\text{and}\qquad X_{t}^{\pi}:=e^{r(T-t)}\bar{X}^{\bar{\pi}}_{t}.
\]
We have
\begin{equation} \label{wealth}
dX_{t}^{\pi}=\pi_{t}\left(\mu(Z_{t})-r\right)dt+\pi_{t}\sigma(Z_{t})dW_{t}^{1}.
\end{equation}

\begin{definition} \label{defi1} A control (or strategy) $\pi=(\pi_{s}, t \leq s \leq T)$ is admissible on the time interval $[t,T]$, written $\pi \in \mathcal{A}_{x,y,z,t}$, if it satisfies the following assumptions:
\begin{enumerate}
   \item[(i)] $\pi$ is progressively measurable;
   \item[(ii)] a unique solution to (\eqref{wealth}) exists and
    \[
      \mathbb{E}^{\eta}_{x,y,z,t} \left[\sup_{t \leq s \leq T} |X_{s}^{\pi}|^{2}\right] < +\infty\qquad \text{for all}\ \eta \in \mathcal{M}.
    \]		
\end{enumerate}
\end{definition}

The investor's objective is to
\begin{equation} \label{problem1}
 \text{minimize} \quad \sup_{\eta \in \mathcal{M}} J^{\pi,\eta}(x,y,z,t)
\end{equation}
over the class of admissible strategies $\mathcal{A}_{x,y,z,t}$, where
\[
J^{\pi,\eta}(x,y,z,t):=\mathbb{E}^{\eta}_{x,y,z,t}\left[-X_{T}^{\pi}-Y_{T}^{\eta}\right].
\]

The problem (\eqref{problem1}) is assumed to be a zero-sum stochastic differential game problem between the market and the investor with controls given by $Q^{\eta}$ and  $\pi$ respectively. We are looking for the value function $V(x,y,z,t)$ and a saddle point i.e. a pair  $(\pi^{*},\eta^{*})\in\mathcal{A}_{x,y,z,t}\times\mathcal{M}$, such that 
\[
 J^{\pi^{*},\eta} (x,y,z,t) \leqslant J^{\pi^{*},\eta^{*}} (x,y,z,t) \leqslant  J^{\pi,\eta^{*}} (x,y,z,t),\quad \forall \pi\in \mathcal{A}_{x,y,z,t}, \ \forall \eta\in\mathcal{M}
\]
and
\[ 
V(x,y,z,t):=J^{\pi^{*},\eta^{*}}(x,y,z,t).
\]

\begin{remark}\label{optimal_pi}
If $(\pi^{*},\eta^{*})\in\mathcal{A}_{x,y,z,t}\times\mathcal{M}$ is a saddle point of problem (\eqref{problem1}), then
\begin{multline*}
\inf_{\pi \in \mathcal{A}_{x,y,z,t}} \sup_{\eta\in\mathcal{M}}J^{\pi,\eta}(x,y,z,t) \leq \sup_{\eta\in\mathcal{M}}J^{\pi^{*},\eta}(x,y,z,t) \leq J^{\pi^{*},\eta^{*}}(x,y,z,t) \\ \leq \inf_{\pi \in \mathcal{A}_{x,y,z,t}} J^{\pi,\eta^{*}} (x,y,z,t)\leq \sup_{\eta\in\mathcal{M}}\inf_{\pi \in \mathcal{A}_{x,y,z,t}} J^{\pi,\eta}(x,y,z,t).
\end{multline*}
In addition, we always have
\[
 \sup_{\eta\in\mathcal{M}}\inf_{\pi \in \mathcal{A}_{x,y,z,t}}J^{\pi,\eta}(x,y,z,t) \leq \inf_{\pi \in \mathcal{A}_{x,y,z,t}} \sup_{\eta\in\mathcal{M}}J^{\pi,\eta}(x,y,z,t).
\]
Summarizing
\[
J^{\pi^{*},\eta^{*}}(x,y,z,t)=\inf_{\pi \in \mathcal{A}_{x,y,z,t}} J^{\pi,\eta^{*}}(x,y,z,t),
\]
so the control $\pi^{*}$ is an optimal financial strategy. 
\end{remark}

For more information about differential games we refer to Fleming and Soner \cite{Fleming} and references therein.

\subsection*{The verification theorem.}\label{sec:4}

The investment problem stated in the  previous section can be solved by applying the stochastic control theory. In this section we establish a link between the Hamilton-Jacobi-Bellman-Isaacs equation and the saddle point of our initial problem. 

Let us recall, that
\begin{equation}\label{model1}
\left\{
\begin{aligned}
dX_{s}^{\pi} &=\pi_{s}(\mu(Z_{s})-r)ds+\pi_{s}\sigma(Z_{s})dW_{s}^{1},  \\
dY_{s}^{\eta}  &=\eta_{s,1}Y_{s}^{\eta} dW_{s}^{1} + \eta_{s,2}Y_{s}^{\eta} dW_{s}^{2},   \\
dZ_{s} &=a(Z_{s}) ds + b(Z_{s})(\rho dW_{s}^{1} + \bar{\rho} dW_{s}^{2}).
\end{aligned}
\right.
\end{equation}
It is convenient to consider $Q^{\eta}$-dynamics of  system (\eqref{model1}). After applying the Girsanov transformation, we have
\begin{equation}\label{model2}
\left\{
\begin{aligned}
 dX_{s}^{\pi} &= \pi_{s}(\mu(Z_{s})-r+\sigma(Z_{s})\eta_{s,1})ds+\pi_{s}\sigma(Z_{s})dW_{s}^{\eta_{1}}, \\
 dY_{s}^{\eta} &= (\eta_{s,1}^{2}+\eta_{s,2}^{2})Y_{s}^{\eta} ds+\eta_{s,1}Y_{s}^{\eta} dW_{s}^{\eta_{1}}+\eta_{s,2}Y_{s}^{\eta} dW_{s}^{\eta_{2}},\\
 dZ_{s} &= (a(Z_{s})+ b(Z_{s})\rho\eta_{s,1}+b(Z_{s})\bar{\rho}\eta_{s,2} )dt + b(Z_{s})(\rho dW_{s}^{\eta_{1}} + \bar{\rho} dW_{s}^{\eta_{2}}),
\end{aligned}
\right.
\end{equation}
where $(W_{s}^{\eta_{1}}, 0\leq s\leq T)$ and $(W_{s}^{\eta_{2}},0\leq s\leq T)$ are $Q^{\eta}$-Brownian motions defined as
\[
\left\{
\begin{aligned}
 dW_{s}^{\eta_{1}} &= dW_{s}^{1}-\eta_{s,1}ds, \\
 dW_{s}^{\eta_{2}} &= dW_{s}^{2}-\eta_{s,2}ds.
\end{aligned}
\right.
\]
Let $\mathcal{L}^{\pi,\eta}$ be a differential operator given by
\begin{align*}
\mathcal{L}^{\pi,\eta}  V (x,y,z,t) :=&  V_{t}+\pi(\mu(z)-r+\sigma(z)\eta_{1})V_{x}+(\eta_{1}^{2}+\eta_{2}^{2})yV_{y} \\
&+ (a(z)+b(z)\rho\eta_{1}+b(z)\bar{\rho}\eta_{2})V_{z}+\frac{1}{2}\pi^{2}\sigma^{2}(z)V_{xx}\\
&+\frac{1}{2}(\eta_{1}^{2}+\eta_{2}^{2})y^{2}V_{yy}+\frac{1}{2}b^{2}(z)V_{zz}+\pi\sigma(z)\eta_{1}yV_{xy}\\
&+\pi\sigma(z)b(z)\rho V_{xz}+b(z)(\rho\eta_{1}+\bar{\rho}\eta_{2})yV_{yz}.
\end{align*}
We can now formulate the Verification Theorem. The proof of this theorem is very similar to the proof of the analogous
theorem from Mataramvura and {\O}ksendal \cite{Mataramvura} (Theorem 3.2) or Zawisza \cite{Zawisza1, Zawisza2} (Theorem 3.1 and Theorem 6.1, respectively), so in this paper we omit it.
\begin{theorem}[Verification Theorem]\label{verification_theorem} Suppose there exists a function 
\[
V \in \mathcal{C}^{2,2,2,1}(\mathbb{R}\times(0,+\infty)\times\mathbb{R} \times [0,T)) \cap \mathcal{C} (\mathbb{R}\times[0,+\infty)\times\mathbb{R} \times [0,T])
\]
and a Markov control 
\[
(\pi^{*},\eta^{*})\in\mathcal{A}_{x,y,z,t}\times\mathcal{M},
\]
such that
\begin{align}
&\mathcal{L}^{\pi^{*}(x,y,z,t),\eta}V(x,y,z,t) \leq 0 \label{first:in1}, \\
&\mathcal{L}^{\pi,\eta^{*}(x,y,z,t)}V(x,y,z,t) \geq 0  \label{second:in1}, \\
&\mathcal{L}^{\pi^{*}(x,y,z,t),\eta^{*}(x,y,z,t)}V(x,y,z,t) = 0\label{third:eq1}, \\
&V(x,y,z,T)=-x-y \label{terminal:cond1}
\end{align}
\flushright for all $\pi \in \mathbb{R}$, $\eta \in \mathbb{R}^{2}$ and  $(x,y,z,t) \in \mathbb{R}\times(0,+\infty)\times\mathbb{R} \times [0,T) $,
\flushleft
and
\begin{equation} \label{uniform}
 \mathbb{E}_{x,y,z,t}^{\eta} \left[ \sup_{t \leq s \leq T} \left|V(X_{s}^{\pi},Y_{s}^{\eta},Z_{s},s)\right|\right] < + \infty 
\end{equation}
\flushright for all  $\pi \in \mathcal{A}_{x,y,z,t}$, $\eta \in \mathcal{M}$ and $ (x,y,z,t) \in  \mathbb{R}\times[0,+\infty)\times\mathbb{R} \times [0,T]$.\flushleft
\medskip

Then
\[
J^{\pi^{*},\eta}(x,y,z,t) \leq V(x,y,z,t) \leq J^{\pi,\eta^{*}}(x,y,z,t),\quad \forall \pi\in \mathcal{A}_{x,y,z,t}, \ \forall \eta\in\mathcal{M}
\]
and
\[ 
V(x,y,z,t)= J^{\pi^{*},\eta^{*}}(x,y,z,t). 
\]
\end{theorem}

\subsection*{Solution to the minimax problem.}\label{sec:5}

To find a saddle point we start with analyzing the Hamilton-Jacobi-Bellman-Isaacs equation
\begin{equation} \label{upper_isaacs}
\min_{\pi \in \mathbb{R}} \max_{\eta \in \mathbb{R}^{2}} \mathcal{L}^{\pi,\eta} V(x,y,z,t)=0,
\end{equation}
i.e.
\begin{align*}
&V_{t}+a(z)V_{z}+\frac{1}{2}b^{2}(z)V_{zz}\\
&+\min_{\pi\in\mathbb{R}}\max_{(\eta_{1},\eta_{2})\in\mathbb{R}^{2}}\biggl\lbrace \pi(\mu(z)-r+\sigma(z)\eta_{1})V_{x}+(\eta_{1}^{2}+\eta_{2}^{2})yV_{y}\\
&+ b(z)(\rho\eta_{1}+\bar{\rho}\eta_{2})V_{z}+\frac{1}{2}\pi^{2}\sigma^{2}(z)V_{xx}+\frac{1}{2}(\eta_{1}^{2}+\eta_{2}^{2})y^{2}V_{yy}\\
&+\pi\sigma(z)\eta_{1}yV_{xy}+\pi\sigma(z)b(z)\rho V_{xz}+b(z)(\rho\eta_{1}+\bar{\rho}\eta_{2})yV_{yz}\biggl\rbrace=0.
\end{align*}
We expect $V(x,y,z,t)$ to be of the form
\begin{equation}\label{given}
V(x,y,z,t)=-x+G(z,t)y,\qquad\text{where}\qquad G(z,T)=-1.
\end{equation}
Then we have
\begin{align*}
yG_{t}+a(z)yG_{z}+\frac{1}{2}b^{2}(z)yG_{zz}&+\min_{\pi\in\mathbb{R}}\max_{(\eta_{1},\eta_{2})\in\mathbb{R}^{2}}\biggl\lbrace-\pi\left(\mu(z)-r+\sigma(z)\eta_{1}\right)\\
&+\left(\eta_{1}^{2}+\eta_{2}^{2}\right)yG+2b(z)(\rho\eta_{1}+\bar{\rho}\eta_{2})yG_{z}\biggl\rbrace=0.
\end{align*}
The maximum over $(\eta_{1},\eta_{2})$ is attained at $( \eta_{1}^{*}(\pi),\eta^{*}_{2})$, where
\begin{align*}
 \eta_{1}^{*}(\pi)&= \frac{\sigma(z)}{2yG(z,t)}\pi-\rho b(z) \frac{G_{z}(z,t)}{G(z,t)},\\
 \eta_{2}^{*}&= -\bar{\rho}b(z)\frac{G_{z}(z,t)}{G(z,t)}.
\end{align*}
For $( \eta_{1}^{*}(\pi),\eta^{*}_{2})$ our equation is of the form
\begin{align}
yG_{t}+a(z)yG_{z}&+\frac{1}{2}b^{2}(z)yG_{zz}+\min_{\pi\in\mathbb{R}}\biggl\lbrace -\pi\left(\mu(z)-r+\sigma(z)\eta^{*}_{1}(\pi)\right)\label{eq2} \\
&+\left((\eta^{*}_{1}(\pi))^{2}+(\eta^{*}_{2})^{2}\right)yG+2b(z)\left(\rho\eta^{*}_{1}(\pi)+\bar{\rho}\eta^{*}_{2}\right)yG_{z}\biggl\rbrace=0.\notag
\end{align}
The minimum over $\pi$ is attained at
\begin{equation}\label{pistar}
\pi^{*}=-2 y G(z,t)\left[\frac{\mu(z)-r}{\sigma^{2}(z)}
-\frac{\rho b(z)}{\sigma(z)}\frac{G_{z}(z,t)}{G(z,t)}\right] .
\end{equation}
It is worth to notice here that 
\begin{equation}\label{eta_pi}
\eta_{1}^{*}(\pi^{*})=-\frac{\mu(z)-r}{\sigma(z)},
\end{equation}
so the saddle point candidate
\begin{equation}\label{saddle_point_candidate}
\left(\pi^{*},(\eta_{1}^{*}(\pi^{*}),\eta_{2}^{*})\right)
\end{equation}
looks as follows
\begin{align*}
\pi^{*}&=-2 y G(z,t)\left[\frac{\lambda(z)}{\sigma(z)}-\frac{\rho b(z)}{\sigma(z)}\frac{G_{z}(z,t)}{G(z,t)}\right]  ,\\
\eta_{1}^{*}(\pi^{*})& =-\lambda(z),\\
\eta_{2}^{*} &= -\bar{\rho}b(z)\frac{G_{z}(z,t)}{G(z,t)},
\end{align*}
where
\[
\lambda(z)=\frac{\mu(z)-r}{\sigma(z)}.
\]
Now we substitute (\eqref{pistar}) into (\eqref{eq2}) and, after dividing by $y$, we get the final equation of the form
\begin{equation}\label{eqresulting}
G_{t}
+\left(a(z)-2\rho b(z)\lambda(z)\right)G_{z}+\frac{1}{2}b^{2}(z)G_{zz}-\bar{\rho}^{2}b^{2}(z)\frac{G^{2}_{z}}{G}+\lambda^{2}(z)G=0,
\end{equation}
 with the boundary condition $G(z,T)=-1$.

\begin{remark}
In Section 3  we rewrite equation (\eqref{eqresulting}) with the boundary condition $G(z,T)=-1$ in a linear form and we provide a set of assumptions which ensure the existence of a classical solution (class $\mathcal{C}^{2,1}$).
\end{remark}

\subsection*{Auxiliary results.}\label{sec:6} 

The following lemmas will be helpful in Section 4 to prove the main theorem and to establish the similarities between our paper and mean-variance optimization methods. In these lemmas we assume that initial conditions $(x,y,r,t)$ are fixed.

\begin{lemma}\label{minimax_equalities_lemma}
Suppose that function $V  \in \mathcal{C}^{2,2,2,1}(\mathbb{R}\times(0,+\infty)\times\mathbb{R} \times [0,T))$ given by (\eqref{given}) is a classical solution to (\eqref{upper_isaacs}). Moreover, let $\left(\pi^{*},(\eta_{1}^{*}(\pi^{*}),\eta_{2}^{*})\right)\in\mathcal{A}_{x,y,z,t}\times\mathcal{M}$ be determined by using (\eqref{saddle_point_candidate}). Then conditions (\eqref{first:in1}) - (\eqref{terminal:cond1}) of Theorem \ref{verification_theorem} are satisfied.
\end{lemma}
\begin{proof}
We already know that 
\[
\max_{\eta \in \mathbb{R}}  \mathcal{L}^{\pi^{*},\eta} V(x,y,z,t) = 0,\qquad \mathcal{L}^{\pi^{*},\eta^{*}} V(x,y,z,t) = 0
\]
and 
\[
  V(x,y,z,T)=-x-y,
\]
which confirms (\eqref{first:in1}), (\eqref{third:eq1}) and (\eqref{terminal:cond1}).

To prove (\eqref{second:in1}) it is sufficient to use (\eqref{eq2}) and (\eqref{eta_pi}) and  verify that 
\[
 \min_{\pi \in \mathbb{R}} \mathcal{L}^{\pi,\eta^{*}(\pi^{*})} V(x,y,z,t)= 0. 
\]
\end{proof}

\begin{lemma} \label{lem_reduction}
Suppose that function $G$ is a classical solution to equation (\eqref{eqresulting}) and  $\left(\pi^{*},(\eta_{1}^{*}(\pi^{*}),\eta_{2}^{*})\right)\in\mathcal{A}_{x,y,z,t}\times\mathcal{M}$ is given by (\eqref{saddle_point_candidate}). Then
\[
 2Y^{\eta^{*}}_{s}G(Z_{s},s)=X^{\pi^{*}}_{s}-x+2yG(z,t), \qquad s \in[t,T].
\]
\end{lemma}
\begin{proof}
It is sufficient to prove only that 
\[
dX^{\pi^{*}}_{s}=d\left(2Y^{\eta^{*}}_{s}G(Z_{s},s)\right).
\]
First of all, note that for the saddle point given by (\eqref{saddle_point_candidate}) system of equations (\eqref{model1}) is of the form
\begin{align}
\begin{split}\label{aa:1}
dX_{s}^{\pi^{*}}=&-2 Y^{\eta^{*}}_{s}G(Z_{s},s)\left[\lambda^{2}(Z_{s})-\rho b(Z_{s})\lambda(Z_{s})\frac{G_{z}(Z_{s},s)}{G(Z_{s},s)}\right]ds\\
&-2 Y^{\eta^{*}}_{s}G(Z_{s},s)\left[\lambda(Z_{s})- \rho b(Z_{s})\frac{G_{z}(Z_{s},s)}{G(Z_{s},s)}\right]dW_{s}^{1}
\end{split}
\end{align}
and
\[
dY^{\eta^{*}}_{s} =-\lambda(Z_{s})Y^{\eta^{*}}_{s}dW_{s}^{1}-\bar{\rho}b(Z_{s})\frac{G_{z}(Z_{s},s)}{G(Z_{s},s)}Y^{\eta^{*}}_{s}dW_{s}^{2}.
\]
Using (\eqref{eqresulting}), we can verify that
\begin{align*}
dG(Z_{s},s)=&\left[2\rho b(Z_{s})\lambda(Z_{s})G_{z}(Z_{s},s)+\bar{\rho}^{2}b^{2}(Z_{s})\frac{G_{z}^{2}(Z_{s},s)}{G(Z_{s},s)}-\lambda^{2}(Z_{s})G(Z_{s},s)\right]ds\\
&+G_{z}(Z_{s},s)b(Z_{s})\left(\rho dW_{s}^{1} + \bar{\rho} dW_{s}^{2}\right).
\end{align*}
Moreover, we have 
\[
d\left(2Y^{\eta^{*}}_{s}G(Z_{s},s)\right)=2G(Z_{s},s)dY^{\eta^{*}}_{s}+2Y^{\eta^{*}}_{s}dG(Z_{s},s)+2dG(Z_{s},s)dY^{\eta^{*}}_{s},
\]
so, by substituting the appropriate dynamics into the above equation, we get the right hand side of (\eqref{aa:1}). 
\end{proof}

\begin{remark} \label{reduction}
Note that the  process  $(Y^{\eta^{*}}_{s},t\leq s\leq T)$ is not directly observable in the financial market, but fortunately  
the above lemma ensures that for the fixed initial conditions $(x,y,z,t)$, rather than the Markov strategy 
\[
\pi^{*}_{s}=-2 Y^{\eta^{*}}_{s}G(Z_{s},s)\left[\frac{\lambda(Z_{s})}{\sigma(Z_{s})}- \frac{\rho b(Z_{s})}{\sigma(Z_{s})}\frac{G_{z}(Z_{s},s)}{G(Z_{s},s)}\right],\qquad s\in[t,T],
\]
we can use
\[
\pi^{*}_{s}=-\left( X^{\pi^{*}}_{s}-x+ 2 yG(z,t)\right)\left[\frac{\lambda(Z_{s})}{\sigma(Z_{s})}- \frac{\rho b(Z_{s})}{\sigma(Z_{s})}\frac{G_{z}(Z_{s},s)}{G(Z_{s},s)}\right],\qquad s\in[t,T].
\]
\end{remark}

\begin{section}{Classical smooth solution to the resulting equation.}\label{sec:7}

To solve equation (\eqref{eqresulting}) with the boundary condition $G(z,T)=-1$ we consider following cases separately:
\\
\\
Case I:$\displaystyle\quad \rho^{2} \neq \frac{1}{2}$
\\
\\
In this case the following ansatz is made (see Zariphopoulou \cite{Zari2})
\[
G(z,t)=-F^{\alpha}(z,t),\qquad\text{where}\qquad F(z,T)=1\qquad\text{and}\qquad \alpha \in\mathbb{R}\backslash\{0\},
\]
to obtain (by dividing by $-\alpha F^{\alpha-1}$)
\[
F_{t}+\left(a(z)-2\rho b(z)\lambda(z)\right)F_{z}+\frac{1}{2}b^{2}(z)F_{zz}+\frac{1}{\alpha}\lambda^{2}(z)F
\]
\[
+\left(\frac{1}{2}(\alpha-1)-\alpha\bar{\rho}^{2}\right)b^{2}(z)\frac{F_{z}^{2}}{F}=0.
\]
Note that for
\[
\alpha=\frac{1}{2\rho^{2}-1},
\]
we have
\begin{equation}\label{eqfinal1}
F_{t}+\left(a(z)-2\rho b(z)\lambda(z)\right)F_{z}+\frac{1}{2}b^{2}(z)F_{zz}+(2\rho^{2}-1)\lambda^{2}(z)F=0.
\end{equation}
\\
Case II:$\displaystyle\quad \rho^{2}=\frac{1}{2}$
\\
\\
In this case if we substitute
\[
G(z,t)=-e^{F(z,t)},\qquad\text{where}\qquad F(z,T)=0,
\]
we get
\begin{equation}\label{eqfinal2}
F_{t}+\left(a(z)-\sqrt{2} b(z)\lambda(z)\right)F_{z}+\frac{1}{2}b^{2}(z)F_{zz}+\lambda^{2}(z)=0.
\end{equation}

Now we give a set of assumptions to ensure the existence of a classical smooth solution to equation (\eqref{eqresulting}) with the boundary condition $G(z,T)=-1$ for any $\rho\in[-1,1]$. 

\begin{remark} \label{important_remark}
From Theorem 1 of  Heath and Schweizer \cite{Heath} it follows that if $a$, $b$, $b \cdot \lambda$, $\lambda^{2}$  are  Lipschitz continuous, $\lambda$ is continuous and bounded and $b^{2}>\epsilon>0$, then there exist the unique classical solutions (class $\mathcal{C}^{2,1}(\mathbb{R} \times [0,T)) \cap\mathcal{C}(\mathbb{R} \times [0,T]) $)  $F_{1}$ and $F_{2}$ to equations (\eqref{eqfinal1}) and (\eqref{eqfinal2}) respectively, that satisfy the Feynman-Kac representations:
\begin{align*}
F_{1}(z,t)&=\mathbb{E}_{z,t}^{\tilde{P}}\left[ \exp\left\{(2\rho^{2}-1)\int_{t}^{T} \lambda^{2}(\tilde{Z}_{s}) ds\right\}\right] ,\\
F_{2}(z,t)&=\mathbb{E}_{z,t}^{\tilde{P}}\left[ \int_{t}^{T} \lambda^{2}(\tilde{Z}_{s}) ds \right],
\end{align*}
where
\[
d\tilde{Z}_{s}=\left[a(\tilde{Z}_{s})-2\rho b(\tilde{Z}_{s})\lambda(\tilde{Z}_{s})\right] ds + b(\tilde{Z}_{s}) d\tilde{W}_{s},\qquad \tilde{Z}_{t}=z
\]
and $(\tilde{W}_{s}, t \leq s \leq T)$ is a Brownian motion with respect to $\tilde{P}$.

Note that, if $\lambda$ is a bounded function, then $F_{1}$ and $F_{2}$ are bounded and $G$ is bounded and bounded away from 0 for any $\rho\in[-1,1]$. 
\end{remark}

\begin{lemma} \label{derivative}
Suppose  $a$, $b$, $b \cdot \lambda$, $\lambda^{2}$ are Lipschitz continuous, $\lambda$ is continuous and bounded, $b^{2}>\epsilon>0$   and $F \in \mathcal{C}^{2,1}(\mathbb{R} \times [0,T)) \cap\mathcal{C}(\mathbb{R} \times [0,T]) $ is a bounded solution to equation  (\eqref{eqfinal1}) or  (\eqref{eqfinal2}). Then the first $z$-derivative of $F$ is bounded.
\end{lemma}

\begin{proof}
To get a bound for $F_{z}$ it is sufficient to estimate the Lipschitz constant. First of all, note that for $z_{1},z_{2} \in (- \infty, c]$ there exists $L_{c}>0$ such that 
\begin{equation}\label{property}
|e^{z_{1}}-e^{z_{2}}| \leq L_{c}|z_{1}-z_{2}|.
\end{equation}
Secondly, for solution to (\eqref{eqfinal1}), using (\eqref{property}) and the fact that $\lambda^{2}$ is Lipschitz continuous and bounded, we obtain the existence of $L>0$ such that
\begin{align*}
|F(z,t)-F(\bar{z},t)| \leq&  L\mathbb{E}^{\tilde{P}}\left[ \int_{t}^{T}\left|\tilde{Z}_{s}(z,t)-\tilde{Z}_{s}(\bar{z},t)\right| ds\right]\\ 
\leq& L T \mathbb{E}^{\tilde{P}} \left[\sup_{t \leq s \leq T} \left|\tilde{Z}_{s}(z,t)-\tilde{Z}_{s}(\bar{z},t)\right|\right],
\end{align*}
where from  notational covenience we wrote $\mathbb{E}^{\tilde{P}}\left[f(\tilde{Z}_{s}(z,t))\right]$ instead of $\mathbb{E}_{z,t}^{\tilde{P}}\left[f(\tilde{Z}_{s})\right]$. Now, using Jensen's inequality and Theorem 1.3.16 from Pham \cite{Pham2}, we have
\[
\exists C_{T}>0\colon\qquad\mathbb{E}^{\tilde{P}} \left[\sup_{t \leq s \leq T} \left|\tilde{Z}_{s}(z,t)-\tilde{Z}_{s}(\bar{z},t)\right|\right] \leq C_{T}|z-\bar{z}|,
\]
which completes the proof in the first case. Naturally, we can get a similar estimate for solution to (\eqref{eqfinal2}).
\end{proof}

\end{section}

\section{Solution to auxiliary optimization problem.}\label{sec:8}
 
In this section we solve portfolio optimization problem (\eqref{problem1}).

\begin{theorem} \label{main2}
Suppose that $a$, $b$, $b \cdot \lambda$, $\lambda^{2}$ are  Lipschitz continuous, $\lambda$ is continuous and bounded, $b$ is bounded and $b^{2}>\epsilon>0$. Then for each initial conditions $(x,y,z,t)$ there exists a Markov saddle point 
\[
\left(\pi^{*},\left(\eta_{1}^{*}(\pi^{*}),\eta_{2}^{*} \right) \right) \in\mathcal{A}_{x,y,z,t}\times\mathcal{M}
\]
for problem (\eqref{problem1}), such that for all $s\in\left[t,T\right]$
\begin{align*}
\pi^{*}_{s}&=-\left( X^{\pi^{*}}_{s}-x+ 2 yG(z,t)\right)\left[\frac{\lambda(Z_{s})}{\sigma(Z_{s})}-\frac{\rho b(Z_{s})}{\sigma(Z_{s})}\frac{G_{z}(Z_{s},s)}{G(Z_{s},s)}\right]  ,\\
\eta_{s,1}^{*}(\pi^{*})& =-\lambda(Z_{s}),\\
\eta_{s,2}^{*} &= -\bar{\rho}b(Z_{s})\frac{G_{z}(Z_{s},s)}{G(Z_{s},s)},
\end{align*}
where $G \in \mathcal{C}^{2,1}(\mathbb{R} \times [0,T)) \cap\mathcal{C}(\mathbb{R} \times [0,T]) $ and is a bounded solution to 
\begin{equation}\label{eqresulting2}
G_{t}
+\left(a(z)-2\rho b(z)\lambda(z)\right)G_{z}+\frac{1}{2}b^{2}(z)G_{zz}-\bar{\rho}^{2}b^{2}(z)\frac{G^{2}_{z}}{G}+\lambda^{2}(z)G=0,
\end{equation}
with the terminal condition $G(z,T)=-1$.
\end{theorem}
\begin{proof}
It follows from Remark \ref{important_remark} and Lemma \ref{derivative} that there exists a classical bounded solution to (\eqref{eqresulting2}), which has bounded derivative $G_{z}$. If we set
\[
 V(x,y,z,t):=-x+G(z,t)y,
\]
then it is sufficient to check whether function $V$ and $\left(\pi^{*},\left(\eta_{1}^{*}(\pi^{*}),\eta_{2}^{*} \right) \right)$ satisfy all conditions of the Verification Theorem. Due to calculations (\eqref{upper_isaacs}) - (\eqref{eqresulting}) and Lemma \ref{minimax_equalities_lemma}, conditions (\eqref{first:in1}) - (\eqref{terminal:cond1}) are fulfilled. Now, we only have to prove that $\left(\pi^{*},\left(\eta_{1}^{*}(\pi^{*}),\eta_{2}^{*} \right) \right)$ belongs to the set $\mathcal{A}_{x,y,z,t}\times\mathcal{M}$ and condition  (\eqref{uniform}) holds:
\\
\\
1. From Lemma \ref{derivative} and Remark \ref{important_remark} we know that the function $G_{z}/G$ is bounded, so $\left(\eta_{1}^{*}(\pi^{*}),\eta_{2}^{*} \right) \in\mathcal{M}$. 
\\
\\
2. Since $G$ is bounded and  $Y^{\eta^{*}}_{s}$ is a solution to the stochastic linear equation with bounded coefficients, using H\"{o}lder's inequality, we have
\[
 \mathbb{E}_{x,y,z,t}^{\eta} \left[ \sup_{t \leq s \leq T} \left|G(Z_{s},s)Y_{s}^{\eta^{*}}\right|\right] \leq \sqrt{\mathbb{E}^{P}\left[\left(\frac{dQ^{\eta}}{dP}\right)^{2}\right]}\cdot\sqrt{\mathbb{E}_{x,y,z,t}^{P}\left[\sup_{t \leq s \leq T} \left|G(Z_{s},s)Y_{s}^{\eta^{*}}\right|^2\right]} < + \infty,
\]
for all $\eta \in \mathcal{M}$.
\\
\\
3. To prove the same with $X_{s}^{\pi^{*}}$ we use the fact that from Lemma \ref{lem_reduction} and Remark \ref{reduction} for the fixed initial conditions $(x,y,z,t)$ the strategy  $\pi^{*}_{s}$ might be rewritten as  
\[
\pi^{*}_{s}=- \left(X^{\pi^{*}}_{s}-x + 2yG(z,t) \right) \left[\frac{\lambda(Z_{s})}{\sigma(Z_{s})}- \frac{\rho b(Z_{s})}{\sigma(Z_{s})}\frac{G_{z}(Z_{s},s)}{G(Z_{s},s)}\right],\qquad s\in\left[t,T\right].
\]
Now, let us define
\[
\zeta(Z_{s},s):= -\left[\frac{\lambda(Z_{s})}{\sigma(Z_{s})}- \frac{\rho b(Z_{s})}{\sigma(Z_{s})}\frac{G_{z}(Z_{s},s)}{G(Z_{s},s)}\right],\qquad s\in\left[t,T\right].
\]
Note  that  $\zeta \cdot (\mu-r)$ and $\zeta \cdot \sigma$ are bounded functions because $\lambda$  and $b$ are bounded. Therefore, with the assistance of  equation (\eqref{aa:1}), we get that  the  process  
\[
K_{s}:= X^{\pi^{*}}_{s}-x + 2yG(z,t),\qquad s\in\left[t,T\right],
\]
is a solution to the following equation
\[
dK_{s} =\zeta(Z_{s},s) (\mu(Z_{s})-r) K_{s}  ds  +  \zeta(Z_{s},s) \sigma(Z_{s}) K_{s} dW_{s}^{1}. 
\]
This is a linear stochastic equation with bounded coefficients, which implies that
\[
\mathbb{E}_{x,y,z,t}^{\eta} \left[ \sup_{t \leq s \leq T} |X_{s}^{\pi^{*}} |\right] \leq \sqrt{\mathbb{E}_{x,y,z,t}^{\eta}  \left[\sup_{t \leq s \leq T} |X_{s}^{\pi^{*}} |^{2}\right]}
\]
\[
\leq\sqrt[4]{\mathbb{E}^{P}\left[\left(\frac{dQ^{\eta}}{dP}\right)^{2}\right]}\cdot\sqrt[4]{\mathbb{E}_{x,y,z,t}^{P}  \left[\sup_{t \leq s \leq T} \left|X_{s}^{\pi^{*}} \right|^{4}\right]} < + \infty, \qquad\forall \eta\in\mathcal{M}.
\]
It means (\eqref{uniform}) is satisfied and confirms the admissibility of $\pi^{*}_{s}$. 
\end{proof}

\section{Mean-variance optimization problem.}\label{sec:9}

Since motivation of our objective function  comes from the mean-variance optimization, it is worth to compare our results with the solution to the latter. To solve the mean-variance optimization problem we consider the following functional
\[
U_{\theta}(X):=\mathbb{E}\left[X\right]-\frac{\theta}{2}\mathbb{E}\left[\left(X-\mathbb{E}\left[X\right]\right)^{2}\right],\qquad X\in\mathcal{L}^{2}(P),
\]
where $\theta>0$ is a risk aversion coefficient and $\mathbb{E}$ denotes the expectation with respect to the measure $P$.

The investor's objective is to
\begin{equation} \label{problem2}
 \text{maximize } \quad \mathcal{I}^{\pi}(x,z,t)
\end{equation}
over the class of admissible\footnote{For more information see Definition \ref{defi1}.}  strategies $\mathcal{A}_{x,z,t}$, where
\[
\mathcal{I}^{\pi}(x,z,t):=\mathbb{E}_{x,z,t}\left[X_{T}^{\pi}\right]-\frac{\theta}{2} \mathbb{E}_{x,z,t}\left[\left(X_{T}^{\pi}-\mathbb{E}_{x,z,t}\left[X_{T}^{\pi}\right]\right)^{2}\right].
\]

We use again the stochastic control methods to obtain a solution. Namely, we can first use the standard Lagrange multipliers technique (see also Zhou and Li \cite{Zhou} for another method resulting
in the same quadratic control problem). Note that
\begin{align}\label{auxiliary22}
\sup_{ \pi \in \mathcal{A}_{x,z,t}}\mathcal{I}^{\pi}(x,z,t) =&\sup_{ \pi \in \mathcal{A}_{x,z,t}} \left\{\mathbb{E}_{x,z,t}\left[X_{T}^{\pi} \right] - \frac{\theta}{2} \mathbb{E}_{x,z,t} \left[\left(X_{T}^{\pi}-\mathbb{E}_{x,z,t}X_{T}^{\pi}\right)^{2}\right]\right\}\notag \\
=&\sup_{A \in \mathbb{R}} \sup_{\pi \in\bar{\mathcal{A}}_{x,z,t}}\left\{ A-\frac{\theta}{2}
\mathbb{E}_{x,z,t}\left[ \left(X_{T}^{\pi}-A \right)^{2}\right]\right\}\notag\\
=&\sup_{A \in \mathbb{R}}\left\{ A-\frac{\theta}{2}\inf_{\pi \in\bar{\mathcal{A}}_{x,z,t}}
\mathbb{E}_{x,z,t}\left[ \left(X_{T}^{\pi}-A \right)^{2}\right]\right\},
\end{align}
 where
\[
\bar{\mathcal{A}}_{x,z,t} = \left\{\pi \in \mathcal{A}_{x,z,t}\colon\ \mathbb{E}_{x,z,t}\left[X_{T}^{\pi}\right]=A\right\},\qquad A\in\mathbb{R}.
\]
In that way we replace the unconstrained mean-variance optimization problem with the constrained maximization of the quadratic objective. Using Lagrange method it is sufficient to minimize the functional
\begin{align} \label{auxiliary}
I^{\pi(\gamma)}(x,z,t)&:=\mathbb{E}_{x,z,t} \left[\left(X_{T}^{\pi(\gamma)}-A\right)^{2}\right] - 2\gamma \mathbb{E}_{x,z,t} \left[X_{T}^{\pi(\gamma)}\right]\notag  \\ 
&=\mathbb{E}_{x,z,t} \left[\left(X_{T}^{\pi(\gamma)}-(A+\gamma) \right)^{2}\right] - 2A \gamma - \gamma^{2}, 
\end{align}
over the class of admissible controls $\pi\in\bar{\mathcal{A}}_{x,z,t}$, determine the solution $\pi^{*}(\gamma)$ and find $\gamma^{*}$, such that 
\[
\mathbb{E}_{x,z,t} \left[X_{T}^{\pi^{*}(\gamma^{*})}\right]=A.
\]
We can use here results from Zawisza \cite{Zawisza2} where minimization of the robust quadratic functional
\[
X_{T}^{\pi} \to \sup_{Q \in \mathcal{Q}}\mathbb{E}_{x,z,t}^{Q}\left[X_{T}^{\pi}-D\right]^{2},\qquad D\in\mathbb{R},
\]
was considered (see also Laurent and Pham \cite{Laurent}). Namely, if we assume that $\mathcal{Q}=\left\{P\right\}$, from Theorem 4.1 in Zawisza \cite{Zawisza2}, we have that the optimal strategy for functional (\eqref{auxiliary}) is given by  
\[
\pi^{*}_{s} (\gamma)= -\left(X_{s}^{\pi^{*} (\gamma)} - (A+ \gamma)\right)\left[ \frac{\lambda(Z_{s})}{\sigma(Z_{s})}+ \frac{\rho b(Z_{s})}{\sigma(Z_{s})} \frac{H_{z}(Z_{s},s)}{H(Z_{s},s)}\right],\qquad s\in[t,T],
\]
where $H$ satisfies 
\[
H_{t}+(a(z)- 2 \rho b(z)\lambda(z)) H_{z}+\frac{1}{2} b^{2}(z) H_{zz} -\rho^{2} b^{2}(z)\frac{H_{z}^{2}}{H}-\lambda^{2}(z)H =0,
\]
with the terminal condition $H(z,T)=1$. It means that $\displaystyle G=-\frac{1}{H}$ is a solution to 
\[
G_{t}+\left(a(z)-2\rho b(z)\lambda(z)\right)G_{z}+\frac{1}{2}b^{2}(z)G_{zz}-\bar{\rho}^{2}b^{2}(z)\frac{G^{2}_{z}}{G}+\lambda^{2}(z)G=0,
\]
where $G(z,T)=-1$. In addition, we have 
\begin{equation} \label{addition}
\pi^{*}_{s} (\gamma)= -\left(X_{s}^{\pi^{*} (\gamma)} - (A+ \gamma)\right)\left[\frac{\lambda(Z_{s})}{\sigma(Z_{s})}- \frac{\rho b(Z_{s})}{\sigma(Z_{s})} \frac{G_{z}(Z_{s},s)}{G(Z_{s},s)}\right],\qquad s\in[t,T],
\end{equation}
which shows that the quadratic optimization is consistent with the monotone optimization with suitable chosen $A$ and $\gamma$ (e.g. Remark \ref{reduction}).

Now we find $\gamma^{*}$, such that 
\[
\mathbb{E}_{x,z,t} \left[X_{T}^{\pi^{*}(\gamma^{*})}\right] = A.
\]
Let us define
\begin{equation}\label{aa2}
P_{s}:=X_{s}^{\pi^{*}(\gamma^{*})}-(A+\gamma^{*}),\qquad s\in[t,T]
\end{equation}
and recall that in the proof of Theorem \ref{main2} we set
\[
\zeta(Z_{s},s)= -\left[\frac{\lambda(Z_{s})}{\sigma(Z_{s})}- \frac{\rho b(Z_{s})}{\sigma(Z_{s})}\frac{G_{z}(Z_{s},s)}{G(Z_{s},s)}\right].
\]
Since $\zeta \cdot (\mu-r)$ and $\zeta \cdot \sigma$ are bounded functions, $P_{s}$ is a solution to the stochastic linear equation with bounded coefficients
\[
dP_{s}=\zeta(Z_{s},s)(\mu(Z_{s})-r)P_{s}ds+\zeta(Z_{s},s)\sigma(Z_{s})P_{s}dW_{s}^{1}.
\]
It means that
\[
P_{s}=\left(x-(A+\gamma^{*})\right)R_{s},
\]
where
\[
R_{s}:=\exp \left\{\int_{t}^{s}\left[\zeta(Z_{u},u)(\mu(Z_{u})-r)-\frac{1}{2}\zeta^{2}(Z_{u},u)\sigma^{2}(Z_{u})\right]du+\int_{t}^{s}\zeta(Z_{u},u)\sigma(Z_{u})\ dW_{u}^{1}\right\}.
\]
Using (\eqref{aa2}), we have
\[
X_{T}^{\pi^{*}(\gamma^{*})}=(x-A)R_{T}+A+\gamma^{*}(1-R_{T}),
\]
and then it is easy to see that
\begin{equation} \label{gamma}
\gamma^{*}(A)=(A-x)\frac{\mathbb{E}_{x,z,t}\left[R_{T}\right]}{1-\mathbb{E}_{x,z,t}\left[R_{T}\right]},\qquad \text{if}\ \ \mathbb{E}_{x,z,t}\left[R_{T}\right]\neq 1.
\end{equation}

Finally, note that
\[
A-\frac{\theta}{2}\mathbb{E}_{x,z,t} \left[\left(X_{T}^{\pi^{*}(\gamma^{*})}-A \right)^{2}\right]=A-\frac{\theta}{2}\mathbb{E}_{x,z,t}\left[\bigl((x-A)R_{T}+\gamma^{*}(A)(1-R_{T})\bigr)^{2}\right],
\]
so the maximum over $A$ is attained at
\begin{equation} \label{A}
A^{*}=x+\frac{1}{\theta}\cdot\frac{1}{\mathbb{E}_{x,z,t}\left[\varphi_{T}^{2}\right]},
\end{equation}
where
\[
\varphi_{T}:=\frac{R_{T}- \mathbb{E}_{x,z,t}\left[R_{T}\right]}{1-\mathbb{E}_{x,z,t}\left[R_{T}\right]}. 
\]
Substituting $\varphi_{T}$ into (\eqref{A}) and $A^{*}$ into (\eqref{gamma}), we get
\[
A^{*}=x+ \frac{1}{ \theta}\cdot \frac{(1-\mathbb{E}_{x,z,t}\left[R_{T}\right])^{2}}{\operatorname{Var}_{x,z,t} \left[R_{T}\right]}  \qquad\text{and} \qquad
\gamma^{*}(A^{*})=\frac{1}{\theta}\cdot \frac{(1-\mathbb{E}_{x,z,t}\left[R_{T}\right])\mathbb{E}_{x,z,t} \left[R_{T}\right]}{\operatorname{Var}_{x,z,t}\left[ R_{T}\right]},
\]
where
\[
\operatorname{Var}_{x,z,t} \left[R_{T}\right]:=\mathbb{E}_{x,z,t}\left[\left(R_{T}-\mathbb{E}_{x,z,t}\left[R_{T}\right]\right)^{2}\right].
\]
Taking into account (\eqref{addition}), we conclude that the mean-variance optimal strategy is given by 
\[
\pi^{*}_{s}= -\left(X_{s}^{\pi^{*}} - x- \frac{1}{\theta}\cdot \frac{1-\mathbb{E}_{x,z,t} \left[R_{T}\right]}{\operatorname{Var}_{x,z,t}\left[ R_{T}\right]}\right)\left[\frac{\lambda(Z_{s})}{\sigma(Z_{s})}- \frac{\rho b(Z_{s})}{\sigma(Z_{s})} \frac{G_{z}(Z_{s},s)}{G(Z_{s},s)}\right],\qquad s\in[t,T], 
\]
whereas the monotone optimal strategy for initial conditions $(x,y,z,t)$ looks as follows
\[
\pi^{*}_{s}=-\left(X_{s}^{\pi^{*}}-x + 2yG(z,t) \right) \left[\frac{\lambda(Z_{s})}{\sigma(Z_{s})}- \frac{\rho b(Z_{s})}{\sigma(Z_{s})}\frac{G_{z}(Z_{s},s)}{G(Z_{s},s)}\right],\qquad s\in[t,T]. 
\]
In the next lemma we prove that both strategies are exactly the same if 
\[
\theta=\frac{1}{2y}.
\]

\begin{lemma} Under conditions of Theorem \ref{main2} for each initial conditions $(x,y,z,t)$ we have 
\[
 \frac{1-\mathbb{E}_{x,z,t} \left[R_{T}\right]}{\operatorname{Var}_{x,z,t} \left[R_{T}\right]}=-G(z,t).
\]
\end{lemma}
\begin{proof}
We start with proving that
\begin{equation}\label{rownosci}
 \mathbb{E}_{x,z,t} \left[R_{T}\right] = \mathbb{E}_{x,z,t}\left[ R_{T}^{2}\right]=H(z,t). 
\end{equation}
First, let us recall that $H$ is a classical solution to
\begin{equation} \label{H}
H_{t}+(a(z)- 2 \rho b(z)\lambda(z)) H_{z}+\frac{1}{2} b^{2}(z) H_{zz} -\rho^{2} b^{2}(z)\frac{H_{z}^{2}}{H}-\lambda^{2}(z)H =0,
\end{equation}
and
\[
R_{T}=\exp \left\{\int_{t}^{T}\left[\zeta(Z_{s},s)(\mu(Z_{s})-r)-\frac{1}{2}\zeta^{2}(Z_{s},s)\sigma^{2}(Z_{s})\right]ds+\int_{t}^{T}\zeta(Z_{s},s)\sigma(Z_{s})\ dW_{s}^{1}\right\},
\]
where 
\[
\zeta(Z_{s},s)= -\left[\frac{\lambda(Z_{s})}{\sigma(Z_{s})}- \frac{\rho b(Z_{s})}{\sigma(Z_{s})}\frac{G_{z}(Z_{s},s)}{G(Z_{s},s)}\right]=  -\left[\frac{\lambda(Z_{s})}{\sigma(Z_{s})}+\frac{\rho b(Z_{s})}{\sigma(Z_{s})}\frac{H_{z}(Z_{s},s)}{H(Z_{s},s)}\right]
\]
and
\[
dZ_{s}=a(Z_{s}) ds + b(Z_{s})(\rho dW_{s}^{1} + \bar{\rho} dW_{s}^{2}),\qquad Z_{t}=z.
\]
Equation (\eqref{H}) can be rewritten into 
\[
H_{t}+ \left(a(z)+\rho b(z)\zeta(z,t)\sigma(z)\right) H_{z}+\frac{1}{2} b^{2}(z) H_{zz}+ \zeta(z,t) \left(\mu(z)-r\right) H =0,
\]
so the solution has the Feynman-Kac representation (e.g. Theorem 1 of  Heath and Schweizer \cite{Heath}) of the following form 
\[
H(z,t)=\mathbb{E}^{\tilde{P}}_{x,z,t} \left[ \exp \left\{\int_{t}^{T}\zeta(Z_{s},s)(\mu(Z_{s})-r)\ ds \right\}\right], 
\]
where
\[
dZ_{s}= \left(a(Z_{s})+\rho b(Z_{s})\zeta(Z_{s},s)\sigma(Z_{s})\right)ds+b(Z_{s})d\tilde{W}_{s},\qquad Z_{t}=z,
\]
$(\tilde{W}_{s},t\leq s\leq T)$ is a Brownian motion with respect to $\tilde{P}$,
\begin{align*}
d\tilde{W}_{s}&=\rho d\tilde{W}_{s}^{1}+\bar{\rho}dW_{s}^{2},\\
d\tilde{W}_{s}^{1}&=dW_{s}^{1}-\zeta(Z_{s},s)\sigma(Z_{s})ds,
\end{align*}
and
\[
\frac{d\tilde{P}}{dP}=\exp \left\{\int_{t}^{T}-\frac{1}{2}\zeta^{2}(Z_{s},s)\sigma^{2}(Z_{s})\ ds 
+\int_{t}^{T}\zeta(Z_{s},s)\sigma(Z_{s})\ dW_{s}^{1}\right\}.
\]
This allows to conclude that 
\[
H(z,t)=\mathbb{E}_{x,z,t} \left[R_{T}\right].
\]

Now, note that
\begin{align*}
R_{T}^{2}=& \exp \biggl\{\int_{t}^{T}\left[ \left(\rho b(Z_{s})\frac{H_{z}(Z_{s},s)}{H(Z_{s},s)}\right)^{2}-\lambda^{2}(Z_{s})\right] ds \\
&-\int_{t}^{T} 2\zeta^{2}(Z_{s},s)\sigma^{2}(Z_{s})\ ds+\int_{t}^{T}2\zeta(Z_{s},s)\sigma(Z_{s})\ dW_{s}^{1}\biggr\}, 
\end{align*}
and equation (\eqref{H}) can be rewritten as 
\[
H_{t}+ \left(a(z)+2\rho b(z)\zeta(z,t)\sigma(z)\right) H_{z}+\frac{1}{2} b^{2}(z) H_{zz}+\left(\rho^{2} b^{2}(z) \frac{H_{z}^{2}}{H^{2}} -\lambda^{2}(z) \right)H =0.
\]
Again the Feynman-Kac representation guarantees the following equality
\[
H(z,t)=\mathbb{E}_{x,z,t}\left[R_{T}^{2}\right].
\]

Finally, let us recall that
\[
G(z,t)=-\frac{1}{H(z,t)},
\]
so using (\eqref{rownosci}), we have 
\[
\frac{1-\mathbb{E}_{x,z,t} \left[R_{T}\right]}{\operatorname{Var}_{x,z,t} \left[R_{T}\right]}=\frac{1}{\mathbb{E}_{x,z,t} \left[R_{T}\right]}\cdot \frac{1-\mathbb{E}_{x,z,t} \left[R_{T}\right]}{1 - \mathbb{E}_{x,z,t}\left[R_{T}\right]}=\frac{1}{H(z,t)}= -G(z,t). 
\]
\end{proof}

\section{General result.}\label{sec:10}

In this section we solve the portfolio optimization problem with performance functional given by (\eqref{functional}).

\begin{theorem}\label{General_result} Under conditions of Theorem \ref{main2} for each initial conditions $(x,y,z,t)$ the stochastic process 
\begin{equation}\label{optymalna_s}
\pi^{*}_{s}=-\left( X^{\pi^{*}}_{s}-x+ 2 yG(z,t)\right)\left[\frac{\lambda(Z_{s})}{\sigma(Z_{s})}- \frac{\rho b(Z_{s})}{\sigma(Z_{s})}\frac{G_{z}(Z_{s},s)}{G(Z_{s},s)}\right],\qquad s\in[t,T],
\end{equation}
is an optimal financial strategy in the portfolio optimization problem with preference criterion given by (\eqref{functional}). Moreover, any optimal solution for the classical mean-variance optimization problem is optimal for functional (\eqref{functional}).
\end{theorem}
\begin{proof}
Let us define
\begin{equation}\label{pomocniczy_f}
\bar{V}_{\theta}(X_{T}^{\pi}):=\inf_{Q\in \bar{\mathcal{Q}}}\left\{\mathbb{E}^{Q}\left[X_{T}^{\pi}\right]+\frac{1}{2\theta}C(Q|P)\right\}.
\end{equation}
Then it is easy to check that
\begin{equation}\label{konkr}
\bar{V}_{\theta}(X_{T}^{\pi})=-\bar{\Lambda}_{\theta}(X_{T}^{\pi})-\frac{1}{2\theta}.
\end{equation}

Secondly, using Theorem 2.1 from Maccheroni et al. \cite{Maccheroni}, we obtain
\begin{equation}\label{nier_1}
 U_{\theta}(X_{T}^{\pi})\leq V_{\theta}(X_{T}^{\pi})\leq\bar{V}_{\theta}(X_{T}^{\pi}),\qquad \forall \pi\in\mathcal{A}_{x,y,z,t},
\end{equation}
so obviously we have
\begin{equation}\label{nier_2}
\sup_{\pi\in\mathcal{A}_{x,y,z,t}} U_{\theta}(X_{T}^{\pi})\leq \sup_{\pi\in\mathcal{A}_{x,y,z,t}} V_{\theta}(X_{T}^{\pi})\leq \sup_{\pi\in\mathcal{A}_{x,y,z,t}}\bar{V}_{\theta}(X_{T}^{\pi}).
\end{equation}
Theorem \eqref{main2} and (\eqref{konkr}) guarantee that for each initial conditions $(x,y,z,t)$ there exists an optimal financial strategy of the form (\eqref{optymalna_s}) in the portfolio optimization problem with the performance functional given by (\eqref{pomocniczy_f}) and from Section 5 we know that it coincides with a solution to classical mean-variance optimization problem. Moreover, it turns out that for $\pi^{*}_{s}$ given by $(\eqref{optymalna_s})$ we have
\[
U_{\theta}(X_{T}^{\pi^{*}})=\bar{V}_{\theta}(X_{T}^{\pi^{*}}),
\]
so taking into account (\eqref{nier_1}) and (\eqref{nier_2}), we get 
\[
\sup_{\pi\in\mathcal{A}_{x,y,z,t}} V_{\theta}(X_{T}^{\pi})=V_{\theta}(X_{T}^{\pi^{*}}). 
\]
To prove the last part of the assertion, let's take any optimal strategy $\hat{\pi}$ for the classical mean-variance optimization problem. Taking into account (\eqref{nier_1}), we have
\[
U_{\theta}(X_{T}^{\pi^{*}})=U_{\theta}(X_{T}^{\hat{\pi}}) \leq \bar{V}_{\theta}(X_{T}^{\hat{\pi}}) \leq \bar{V}_{\theta}(X_{T}^{\pi^{*}})=U_{\theta}(X_{T}^{\pi^{*}}).
\]
Therefore,
\[
\bar{V}_{\theta}(X_{T}^{\hat{\pi}}) = \bar{V}_{\theta}(X_{T}^{\pi^{*}}).
\]

\end{proof}

\section{Explicit solution to the square root factor process.}\label{sec:11} In the previous sections we developed the theory well enough to get more insight in the Markowitz optimal portfolio in the stochastic factor models. In this part, we consider some particular factor market models which admit an explicit solution to the mean-variance problem. The fundamental example of such a family of models are Heston stochastic volatility models. This kind of models usually do not satisfy the boundedness condition which was assumed in the previous sections and therefore the verification result needs a separate justification. Solutions obtained in this section can be compared further to the standard Black-Scholes model. We should also mention here that solutions to models considered in this section were obtained earlier by Shen and Zeng \cite{Shen}, even in slightly more general jump-diffusion framework. However, they used BSDE methods and their formula is not as explicit as ours.  

The model, which we consider in this section, has the following form
\begin{equation}\label{model_explicit1}
\left\{
\begin{aligned}
dS_{t} &=[r+ \lambda Z_{t} \sigma(Z_{t})] S_{t}  dt +  \sigma(Z_{t}) \sqrt{Z_{t}} S_{t}  dW_{t}^{1},   \\
dZ_{t} &= \kappa (\xi - Z_{t}) dt + b \sqrt{Z_{t}}(\rho dW_{t}^{1} + \bar{\rho} dW_{t}^{2}), \qquad 2\kappa \xi > b^{2}, \quad \kappa >0.  
\end{aligned}
\right.
\end{equation}
We avoid here the dynamics for the bank account $B_{t}$ because we work with $T$-forward values in our framework. Apart from the unboundedness of the coefficients, the second difference here is that the factor process $Z_{t}$ takes only positive values. Therefore, when considering the HJB equation, we have to limit ourselves to the space $\mathbb{R} \times \mathbb{R}^{+} \times [0,T]$. In spite of this fact, we can easily follow the path in Section 5 to find the optimal portfolio candidate
\[
\pi^{*}_{s}=-\left( X^{\pi^{*}}_{s}-x-  \frac{1}{\theta}\frac{1}{H(z,t)}\right)\left[\frac{\lambda}{\sigma(Z_{s})}-\frac{\rho b}{\sigma(Z_{s})}\frac{H_{z}(Z_{s},s)}{H(Z_{s},s)}\right],\qquad s\in[t,T],
\]
where $H(z,t)$ is a solution to the resulting equation of the form
\begin{equation}\label{eqresulting3}
H_{t}
+\left[\kappa(\xi-z) -2\rho b  \lambda z\right]H_{z}+\frac{1}{2}b^{2} zH_{zz}-\rho^{2}b^{2} z\frac{H^{2}_{z}}{H} - \lambda^{2} zH=0,\qquad z \geq 0,
\end{equation} 
with the terminal condition $H(z,T)=1$. 

Following Zeng and Taksar \cite{Taksar}, we assume that
\[
H(z,t)=  e^{A(t) z+ B(t)}. 
\]
Substituting the above ansatz into equation (\eqref{eqresulting3}), we get
\begin{gather}
A'(t)+\left(\frac{1}{2} b^{2} - \rho^{2} b^{2}\right) A^{2}(t) + \left(- \kappa - 2 \rho b \lambda\right) A(t) - \lambda^{2} =0, \nonumber \\ 
B'(t) + \kappa \xi A(t)=0.  \nonumber
\end{gather}
In order to solve these equations we consider the quadratic equation of the form
\begin{equation} \label{quadratic}
 \left(\frac{1}{2} b^{2} - \rho^{2} b^{2}\right)x^{2} + \left(- \kappa - 2 \rho b \lambda\right) x - \lambda^{2}=0,
\end{equation}
with the following discriminant
\[
\Delta= \kappa^{2}+4 \kappa \rho \lambda b + 2 \lambda^{2} b^{2}. 
\]
To avoid singularities and exceptions, we always assume that $\Delta>0$ and $\rho^{2} \neq \frac{1}{2}$. In the other cases we have to consider further specification with respect to the time horizon $T>0$. The complete analysis is presented in Proposition 3.1, Zeng and Taksar \cite{Taksar}. The mentioned paper is dedicated to the HARA utility objective with the risk aversion coefficient $\alpha$. Surprisingly, if we take Zeng and Taksar's resulting equation, substitute $\alpha=2$ and apply their results, then we get tools perfectly suited to our case. 

Under our conditions equation (\eqref{quadratic}) has the following two solutions
\[
y_1=\frac{ \kappa + 2 \rho b \lambda + \sqrt{\Delta}}{b^{2} - 2\rho^{2} b^{2}}\qquad\text{and}\qquad y_2=\frac{ \kappa+ 2 \rho b \lambda-\sqrt{\Delta}}{  b^{2} - 2\rho^{2} b^{2}},
\]
so
\begin{equation}\label{A(t)}
A(t)= \frac{e^{p(y_{1}-y_{2})(T-t)} -1}{e^{p(y_{1}-y_{2})(T-t)} -\frac{y_{2}}{y_{1}}} y_{2},\qquad\text{where}\quad p=\frac{1}{2}b^{2} - \rho^{2} b^{2} 
\end{equation}
and
\[
B(t)= \kappa \xi \int_{t}^{T}A(s) ds.
\]
For further applications we need to have the following lemma.
\begin{lemma}\label{L_A(t)} Function $A(t)$ given by (\eqref{A(t)}) is less than or equal to 0 for all $t \in [0,T]$.
\end{lemma}
\begin{proof} Note that $A(t)$ can be rewritten as follows
\[
A(t)= \frac{e^{p(y_{1}-y_{2})(T-t)} -1}{y_{1}e^{p(y_{1}-y_{2})(T-t)} -y_{2}}y_{1} y_{2}.
\]
Moreover, we always have $e^{p(y_{1}-y_{2})(T-t)}= e^{\sqrt{\Delta}(T-t)} >1$ and from Vieta's formulas, we get
\[
y_{1}y_{2} = - \frac{\lambda^{2}}{p}.
\]
Case I: $y_{1}y_{2}=0$ implies $A(t)=0$. 
\\
\\
Case II:$\quad y_{1}y_{2}<0\ (p > 0)$
\\
\\
In this case $y_{1}>0>y_{2}$ and the sign of $A(t)$ is inherited from $y_{1}y_{2}$. 
\\
\\
Case III:$\quad y_{1}y_{2}>0\ (p < 0)$
\\
\\
In this case $2 \rho^{2} >1$ and at the same time $\kappa+ 2 \rho b \lambda-\sqrt{\Delta} >0$, which implies that $y_{1} < y_{2} <0$ and $\frac{y_{2}}{y_{1}} <1$. Formula (\eqref{A(t)}) completes the reasoning.
\end{proof}
To formally prove that the strategy $\pi^{*}$ is optimal, we should extend the class of optimal strategies. 

\begin{definition}\label{defi2} A control (or strategy) $\pi=(\pi_{s}, t \leq s \leq T)$ is admissible on the time interval $[t,T]$, written $\pi \in \mathcal{\bar{A}}_{x,z,t}$, if it satisfies the following assumptions:
\begin{enumerate}
   \item[(i)] $\pi$ is progressively measurable;
   \item[(ii)] there exists a unique solution to the following equation
    \[
    dX_{s}^{\pi}=\pi_{s}\lambda Z_{s} \sigma(Z_{s})ds+\pi_{s}\sigma(Z_{s})\sqrt{Z_{s}}dW_{s}^{1},\qquad s\in[t,T];
		\] 
   \item[(iii)] there exists a localizing sequence of stopping times  $(\tau_{n}, n \in \mathbb{N})$, such that
    \[
    H(Z_{T \wedge \tau_{n}},T \wedge \tau_{n})\left[X_{T \wedge \tau_{n}}^{\pi}\right]^{2}
    \] 	
is uniformly integrable and
\[
\mathbb{E}_{x,z,t} \left[\int_{t}^{T\wedge \tau_{n}}\left[(X_{s}^{\pi})^{4} +1 \right][\sigma(Z_{s})+1]Z_{s} ds\right] < +\infty,
\]
where
\[
H(z,t)=e^{A(t) z + B(t)}.
\] 
\end{enumerate} 
\end{definition}
Note that if
\[
\mathbb{E}_{x,z,t} \left[\sup_{t \leq s \leq T}[X_{s}^{\pi}]^{2}\right] < +\infty,
\]
then the above conditions are satisfied, because $\tau_{n}$ can be set as the first exit time from a sequence of open and bounded subsets exhausting the set $\mathbb{R} \times \mathbb{R}^{+}$. 

\begin{theorem}\label{Heston} In model (\eqref{model_explicit1}), under conditions $\Delta>0$, $\frac{1}{2}b^{2} - \rho^{2} b^{2} \neq 0$ and for each fixed initial conditions  $(x,z,t)$, the stochastic process
\begin{align*}
\pi^{*}_{s}&=-\left(X_{s}^{\pi^{*}}-x-\frac{1}{\theta} e^{-A(t)z -B(t)}\right)\left[\frac{\lambda}{\sigma(Z_{s})}-\frac{\rho b A(s)}{\sigma(Z_{s})}\right],\qquad s\in[t,T],
\end{align*}
is an optimal financial strategy for the classical mean-variance  functional \eqref{problem2}. 
\end{theorem}

\begin{proof}
Keeping in mind the proof of Theorem 4.1 in Zawisza \cite{Zawisza2} and calculations done in Section 5, it is sufficient to prove that for all $D \in \mathbb{R}$ the strategy 
\[
\pi_{D}(x,z,s) = -(x-D)\left[\frac{\lambda}{\sigma(z)}-\frac{\rho b A(s)}{\sigma(z)}\right],
\]
is optimal for the problem 
\[
\mathbb{E}_{x,z,t}(X_{T}^{\pi} - D)^{2} \rightarrow \min_{\pi \in \mathcal{\bar{A}}_{x,z,t}}. 
\]

To verify that, we  can follow the standard proof of the verification theorem. Namely, first let's take any admissible strategy $\pi$ and apply the It\^{o} formula to the function 
\[
V(x,z,t)=(x-D)^{2} H(z,t)
\]
and the process $X^{\pi}_{s}$ to obtain
\[
\mathbb{E}_{x,z,t} V(X^{\pi}_{T \wedge \tau_{n}}, Z_{T \wedge \tau_{n}},T \wedge \tau_{n}) \leq \mathbb{E}_{x,z,t} (X^{\pi}_{T \wedge \tau_{n}}-D)^{2},
\]
where $(\tau_{n}, n \in \mathbb{N})$ is the localizing sequence of stopping times from Definition \ref{defi2}. Then, we can pass to the limit under the expected value by taking the advantage from the uniform integrability condition for the family
\[
H(Z_{T \wedge \tau_{n}},T \wedge \tau_{n})\left[X_{T \wedge \tau_{n}}^{\pi}\right]^{2}
\] 	
and the fact that $A(t)\leq 0$ (see Lemma \ref{L_A(t)}), $z\geq 0$ and consequently $H(z,t) \leq 1$. 

Now, we take into consideration the strategy $\pi_{D}^{*}$. Applying the It\^o formula, we obtain 
\[
\mathbb{E}_{x,z,t} V(X_{T \wedge \tau_{n}}^{\pi_{D}^{*}}, Z_{T \wedge \tau_{n}},T \wedge \tau_{n}) = V(x,z,t),
\]
where 
\[
\tau_{n}:= \inf \left\{s \geq t:  Z_{s} \notin \mathcal{O}_{n}\right\}
\]
and $\{\mathcal{O}_{n}\}_{n \in \mathbb{N}}$ is an increasing family of open and bounded sets exhausting $\mathbb{R}^{+}$.

Note that 
\[
X^{\pi^{*}_{D}}_{s}=(x-D) R_{s} +D \quad\text{and}\quad   V(X_{T \wedge \tau_{n}}^{\pi_{D}^{*}}, Z_{T \wedge \tau_{n}},T \wedge \tau_{n}) = (x-D)^{2}R_{T \wedge \tau_{n}}^{2} H( Z_{T \wedge \tau_{n}},T \wedge \tau_{n}),
\]
where 
\[
R_{s}=\exp \biggl\{\int_{t}^{s}\lambda \sigma(Z_{u}) \zeta(u) Z_{u}-\frac{1}{2}\zeta^{2}(u)\sigma^{2}(Z_{u}) Z_{u}\ du \\
+\int_{t}^{s}\zeta(u)\sigma(Z_{u}) \sqrt{Z_{u}}\ dW_{u}^{1}\biggr\},
\]
\[
R_{s}^{2}=\exp \biggl\{\int_{t}^{s}\left[\lambda \sigma(Z_{u}) \zeta(u) + \zeta^{2}(u)\sigma^{2}(Z_{u})\right] Z_{u} du -2\int_{t}^{s}\zeta^{2}(u)\sigma^{2}(Z_{u}) Z_{u}\ du \\
+\int_{t}^{s}2\zeta(u)\sigma(Z_{u}) \sqrt{Z_{u}}\ dW_{u}^{1}\biggr\}
\]
and 
\[
\zeta(u) = - \left[\frac{\lambda}{\sigma(Z_{u})}-\frac{\rho b A(u)}{\sigma(Z_{u})}\right].
\]

Now, it is sufficient to apply the proof of Proposition A1 form Zeng and Taksar \cite{Taksar} (with $\alpha=2$) to prove that the family $R_{T \wedge \tau_{n}}^{2}H(Z_{T \wedge \tau_{n}},T \wedge \tau_{n})$ is uniformly integrable. Then, we are able to pass to the limit ($n \to +\infty$), to get
\[
\mathbb{E}_{x,z,t}(X_{T}^{\pi_{D}^{*}} -D)^{2}=\mathbb{E}_{x,z,t}V(X_{T}^{\pi_{D}^{*}}, Z_{T},T )= V(x,z,t).
\]
\end{proof}

\begin{theorem}
Under conditions of Theorem \ref{Heston}, every Markowitz optimal strategy (in the class $\mathcal{\bar{A}}_{x,z,t}$) is also optimal 
for monotone functional (\eqref{functional}).
\end{theorem}

\begin{proof}
We can deduce from \eqref{nier_2} that it is sufficient for us to verify that
\begin{align*}
\pi^{*}_{s}&=-\left(X_{s}^{\pi^{*}}-x-\frac{1}{\theta} e^{-A(t)z -B(t)}\right)\left[\frac{\lambda}{\sigma(Z_{s})}-\frac{\rho b A(s)}{\sigma(Z_{s})}\right],\qquad s\in[t,T],
\end{align*}
is optimal for functional (\eqref{pomocniczy_f}), when the infimum is taken under only one measure $Q^{\eta^{*}}$. Therefore, we will prove that $\pi^{*}_{s}$ is the minimum for the functional
\[
J^{\pi}(x,y,z,t):=\mathbb{E}^{\eta^{*}}_{x,y,z,t}\left[-X_{T}^{\pi}-Y_{T}^{\eta^{*}}\right],
\]
where $\eta^{*}$ is determined by 
\begin{align*}
\eta_{1}^{*}& =-\lambda(z),\\
\eta_{2}^{*} &= -\bar{\rho}b(z)\frac{G_{z}(z,t)}{G(z,t)} \quad\text{and}\quad G(z,t)= - \frac{1}{H(z,t)}.
\end{align*}
By applying the It\^{o} formula (like in the proof of Lemma \ref{lem_reduction}), we can prove
\[
2Y^{\eta^{*}}_{s}G(Z_{s},s)=X^{\pi^{*}}_{s}-x+2yG(z,t), \qquad s \in[t,T].
\]
Uniform integrability condition for $H(Z_{T \wedge \tau_{n}},T\wedge \tau_{n}) [X_{T \wedge \tau_{n}}^{\pi^{*}}]^{2}$ ensures that $Y^{\eta^{*}}_{s}$ is a square integrable continuous martingale and consequently an uniformly integrable martingale.

Now, let us define $V(x,y,z,t)=-x + G(z,t)y$. We can easily verify, by direct differentiation, that the suitable HJB equation is satisfied.   

We prove first the following inequality
\begin{equation}\label{veri}
V(x,y,z,t) \leq J^{\pi}(x,y,z,t), \quad \pi \in \mathcal{\bar{A}}_{x,z,t}.
\end{equation}
We can take an admissible strategy $\pi$ and apply the It\^{o} rule, in order to get
\begin{align*}
V(x,y,z,t) &\leq \mathbb{E}^{\eta^{*}}_{x,y,z,t} V(X^{\pi}_{T \wedge \tau_{n}}, Z_{T \wedge \tau_{n}},T \wedge \tau_{n})=\mathbb{E}^{\eta^{*}}_{x,y,z,t}\left[-X^{\pi}_{T \wedge \tau_{n}}+G(Z_{T \wedge \tau_{n}},T \wedge \tau_{n})Y_{T \wedge \tau_{n}}^{\eta^{*}}\right]\\ & \quad = \mathbb{E}_{x,y,z,t}\left[-\frac{1}{y} Y_{T \wedge \tau_{n}}^{\eta^{*}}X_{T \wedge \tau_{n}}^{\pi}+ \frac{1}{y}G(Z_{T \wedge \tau_{n}},T \wedge \tau_{n})[Y_{T \wedge \tau_{n}}^{\eta^{*}}]^{2}\right],
\end{align*}
where $(\tau_{n}, n \in \mathbb{N})$ is a localizing sequence of stopping times from Definition \ref{defi2}. The last equality is implied by the relation 
$y\frac{dQ^{\eta^{*}}}{dP}=Y_{T}^{\eta^{*}}$ and the martingality of the process $Y_{t}^{\eta^{*}}$.

The maximization of the quadratic function $u_{1}a+u_{2}a^{2}$  with respect to $a$ yields
\[
u_{1}a+u_{2}a^{2} \leq -\frac{u_{1}^{2}}{4u_{2}}, \qquad u_{1},a \in \mathbb{R}\ \text{and}\ u_{2}<0.
\]
Substituting $u_{1}=-\frac{1}{y} X_{T \wedge \tau_{n}}^{\pi}$, $u_{2}=\frac{1}{y}G(Z_{T \wedge \tau_{n}},T \wedge \tau_{n})$ and $a=Y_{T \wedge \tau_{n}}^{\eta^{*}}$, we get
\[
-\frac{1}{y} Y_{T \wedge \tau_{n}}^{\eta^{*}}X_{T \wedge \tau_{n}}^{\pi}+ \frac{1}{y}G(Z_{T \wedge \tau_{n}},T \wedge \tau_{n})[Y_{T \wedge \tau_{n}}^{\eta^{*}}]^{2} \leq \frac{1}{4y} H(Z_{T \wedge \tau_{n}},T \wedge \tau_{n}) [X_{T \wedge \tau_{n}}^{\pi}]^{2}.
\]
The strategy $\pi$ is admissible, so the family 
\[
\frac{1}{4y} H(Z_{T \wedge \tau_{n}},T \wedge \tau_{n}) [X_{T \wedge \tau_{n}}^{\pi}]^{2},
\]
is uniformly integrable.  
To assert formula \eqref{veri} it is  sufficient to apply the Fatou Lemma to the following reverse inequality
\[
-V(x,y,z,t) \geq \mathbb{E}_{x,y,z,t}\left[\frac{1}{y} Y_{T \wedge \tau_{n}}^{\eta^{*}}X_{T \wedge \tau_{n}}^{\pi} -\frac{1}{y}G(Z_{T \wedge \tau_{n}},T \wedge \tau_{n})[Y_{T \wedge \tau_{n}}^{\eta^{*}}]^{2}\right].
\]
The Fatou Lemma is possible because the integrand on the right hand side is bounded below by  the uniformly integrable sequence.

%

To finish the proof, we should only verify that
\[
V(x,y,z,t)= J^{\pi^{*}}(x,y,z,t)= \mathbb{E}^{\eta^{*}}_{x,y,z,t}\left[-X_{T}^{\pi^{*}}+Y_{T}^{\eta^{*}}\right].
\]
Here, we just need to employ the It\^{o} formula
\begin{align*}
V(x,y,z,t) &= \mathbb{E}^{\eta^{*}}_{x,y,z,t} V(X^{\pi^{*}}_{T \wedge \tau_{n}}, Z_{T \wedge \tau_{n}},T \wedge \tau_{n})=\mathbb{E}^{\eta^{*}}_{x,y,z,t}\left[-X^{\pi^{*}}_{T \wedge \tau_{n}}+G(Z_{T \wedge \tau_{n}},T \wedge \tau_{n})Y_{T \wedge \tau_{n}}^{\eta^{*}}\right]\\ & \quad = \mathbb{E}_{x,y,z,t}\left[-\frac{1}{y} Y_{T \wedge \tau_{n}}^{\eta^{*}}X_{T \wedge \tau_{n}}^{\pi^{*}}+ \frac{1}{y}G(Z_{T \wedge \tau_{n}},T \wedge \tau_{n})[Y_{T \wedge \tau_{n}}^{\eta^{*}}]^{2}\right],
\end{align*}
where $(\tau_{n}, n \in \mathbb{N})$ is a suitable localizing sequence of stopping times.
Now, we use the following duality
\[
Y^{\eta^{*}}_{s}=\frac{1}{2} \frac{1}{G(Z_{s},s)}  X^{\pi^{*}}_{s}+ \frac{1}{2}\frac{1}{G(Z_{s},s)}[-x+2yG(z,t)]=-\frac{1}{2} H(Z_{s},s) X^{\pi^{*}}_{s}- \frac{1}{2}H(Z_{s},s)[-x+2yG(z,t)]
\]
and pass to the limit ($n \to +\infty$), using the uniform integrability condition for three families
 \[
H(Z_{T \wedge \tau_{n}},T \wedge \tau_{n}) [X_{T \wedge \tau_{n}}^{\pi^{*}}]^{2}, \quad H^{2}(Z_{T \wedge \tau_{n}},T \wedge \tau_{n}) [X_{T \wedge \tau_{n}}^{\pi^{*}}]^{2}\quad\text{and}\quad H(Z_{T \wedge \tau_{n}},T \wedge \tau_{n}) X_{T \wedge \tau_{n}}^{\pi^{*}}.
\] 
The first one has already been commented in the proof of Theorem \ref{Heston}, the rest are its implications. Namely, the second one is implied by the condition  $H(z,t) \leq 1$ and the last one can be proved by the application of the Cauchy-Schwartz inequality.
\end{proof}

\bigskip 

Now, we would like to consider the solution with the emphasis on the following two special cases:
\\
\\
Case I: The volatility coefficient $\sigma(z)=\bar{\sigma}$ is constant. Then
\[
\pi^{*}_{s}=-\left( X_{s}^{\pi^{*}}-x-\frac{1}{\theta} e^{-A(t)z -B(t)}\right)\left[\frac{\lambda}{\bar{\sigma}}-\frac{\rho b A(s)}{\bar{\sigma}}\right],\qquad s\in[t,T].
\]
Case II: The stochastic factor affects only the excess return but not the volatility i.e. $\sigma(z)=\frac{\bar{\sigma}}{\sqrt{z}}$. Then
\[
\pi^{*}_{s}=-\left( X_{s}^{\pi^{*}}-x-\frac{1}{\theta} e^{-A(t)z -B(t)}\right)\left[\frac{\lambda}{\bar{\sigma}}-\frac{\rho b A(s)}{\bar{\sigma}}\right]\sqrt{Z_{s}}, \qquad s\in[t,T].
\]
It is also worth to compare the above cases to the standard Black-Scholes market
\[
dS_{t} =[r+ \lambda  \bar{\sigma}] S_{t} dt + \bar{\sigma}S_{t} dW_{t}^{1},
\]
where
\[
\pi^{*}_{s}=-\left( X_{s}^{\pi^{*}}- x- \frac{1}{\theta} e^{\lambda^{2}(T-t)} \right)\frac{\lambda }{\bar{\sigma}},\qquad s\in[t,T].
\]
In all three cases the absolute value invested in $S_{t}$ is determined in the following way
\[
\pi^{*}_{s}=(X_{s}^{\pi^{*}}-D(x,z,t)) P(Z_{s},s),
\]
so is proportional to the excess of the current wealth $X_{s}^{\pi^{*}}$ over some target level determined by $D(x,z,t)$. The term $P(z,s)$ can be interpreted as the proportion value. We can see that the presence of the stochastic factor modulates the target level $D(x,z,t)$, but important are only the coefficients of the factor dynamics and the value of the stochastic factor at the beginning of the investing period. The proportion $P(z,s)$ depends on the coefficients of the factor dynamics and when considering the model presented in Case II, on the current level of the stochastic factor as well. The direction and magnitude of the factor impact depends on many different configurations of the factor current value and  model coefficients.

\section{Conclusion.}\label{sec:12}

We examined the continuous time portfolio optimization problem the stochastic factor model assuming that the preference criterion is the monotone mean-variance functional introduced by  Maccheroni et al. \cite{Maccheroni}. We have solved the problem under general conditions and have proved that the the optimal mean-variance strategy is optimal also for the monotone criterion. This shows that continuous time  mean-variance investor behaves like a monotone investor, even in economically important incomplete market models. It would be also interesting to check this property in other class of models, for example in jump-diffusion models.  
%
%
%

\section*{Acknowledgments.}
We would like to express our sincere gratitude to the Referees for a list of helpful remarks and a careful reading of the manuscript.


\bibliographystyle{ormsv080} 

\begin{thebibliography}{HD}
\normalsize

\baselineskip=17pt

\bibitem{Artzner}  P. Artzner,  F. Delbaen , JM. Eber, D. Heath \emph{Coherent measures of risk}, Math. Finance 9 (1999),   203 -- 228.

\bibitem{Basak} S. Basak, G. Chabakauri, \emph{Dynamic mean-variance asset allocation}, Review of Financial Studies, 23 (2010), 2970 -- 3016.

\bibitem{Bauerle} N. Bauerle, S. Grether, \emph{Complete markets do not allow free cash 
ow streams}, Math. Meth. Oper.
Res. 81 (2015) 137 -- 146.


\bibitem{Bordigioni} G. Bordigoni,  A. Matoussi, M. Schweizer, \emph{A stochastic control approach to a robust utility maximization problem}, Stochastic Analysis and Applications. The Abel Symposium 2005.
Springer 2007, 125 -- 151.

\bibitem{Campbell}  J.Y. Campbell, L.M.  Viceira  \emph{Consumption and portfolio decisions when expected returns are time
varying}, Q. J. Econ. 114 (1999) 433 -- 495.

\bibitem{Cui}  X. Cui, D. Li, S. Wang, S. Zhu, \emph{Better than dynamic mean variance: Time inconsistency and free cash flow stream}, Math. Finance 22 (2012), 345 -- 378.

\bibitem{Elliott4} R. J. Elliott, T. K. Siu, \emph{A BSDE approach to a risk-based optimal investment of an insurer}, Automatica J. IFAC, 47 (2011), 253 -- 261. 

\bibitem{Elliott} R. J. Elliott, T. K. Siu, \emph{On risk minimizing portfolios under a Markovian regime-switching Black-Scholes economy}, Ann. Oper. Res. 176 (2010), 271 -- 291.

\bibitem{Elliott3} R. J. Elliott, T. K. Siu, \emph{Portfolio risk minimization and differential games}, Nonlinear Anal. 71 (2009), 2127 -- 2135. 

\bibitem{Elliott5} R. J. Elliott, T. K. Siu, \emph{Risk-based indifference pricing under a stochastic volatility model}, Commun. Stoch. Anal. 4 (2010), 51 -- 73.

\bibitem{Fleming2} WH. Fleming, D.  Hern\'{a}ndez-Hern\'{a}ndez,   \emph{An optimal consumption model with stochastic volatility},
Finance Stoch. 7 (2003), 245 -- 262.


\bibitem{Fleming} W. Fleming, H. M. Soner, \emph{Controlled Markov Processes and Viscosity Solutions}, 2nd Edition, Springer, New York, 2006.

\bibitem{Follmer} H. F\"{o}llmer, A. Schied, \emph{Convex measures of risk and trading constraints}, Finance and Stochastics, 6 (2012), 429 -- 447.

\bibitem{Frittelli} M. Frittelli, E. Rosazza-Gianin, \emph{Putting order in risk measures}, Journal of
Banking and Finance, 26 (2002), 1473 -- 1486.

\bibitem{Gilboa} I. Gilboa, D. Schmeidler  \emph{Maxmin expected utility with non-unique prior}, J. Math. Econ. 18 (1989) 141--153.

\bibitem{Heath}  D. Heath, M. Schweizer, \emph{Martingales versus PDEs in finance: an equivalence result with examples}, J. Appl. Probab. 37 (2000), 947 -- 957.

\bibitem{Schied} D. Hern\'{a}ndez, A. Schied, \emph{A control approach to robust utility maximization with logarithmic utility and time-consistent penalties}, Stochastic Process. Appl. 117 (2007), 980 -- 1000.

\bibitem{Kallsen} J. Kallsen, J. Muhle-Karbe, N. Shenkman, R. Vierthauer \emph{Discrete time variance-optimal hedging in
affine stochastic volatility models}, R. Kiesel, M Scherer, eds. Alternative Investments and Strategies (World
Scientific, Singapore), 375--393.

\bibitem{Kim} TS. Kim, E. Omberg \emph{Dynamic nonmyopic portfolio behavior}, Rev. Financ. Stud. 9 (1996), 141 -- 161.

\bibitem{Kraft} H. Kraft, T. Seiferling, FT. Seifried, \emph{Optimal consumption and investment with epstein - zin
recursive utility}. Finance Stoch. 21 (2017), 187 -- 226.

\bibitem{Laurent} J. Laurent,  H. Pham, \emph{Dynamic programming and mean-variance hedging}, Finance Stoch. 3 (1999), 83 -- 110.

\bibitem{Li} D. Li, W. Ng,  \emph{Optimal dynamic portfolio selection: multiperiod mean-variance formulation}, Math.
Finance, 10 (2000), 387 -- 406.

\bibitem{Liu} J. Liu, \emph{Portfolio selection in stochastic environments}, Rev. Financ. Stud. 20 (2007), 1 -- 39.

\bibitem{Macheroni2} F. Maccheroni, M. Marinacci, A. Rustichini,  \emph{Ambiguity aversion, robustness, and the variational
representation of preferences}, Econometrica 74 (2006), 1447 -- 1498.

\bibitem{Maccheroni} F. Maccheroni, M. Marinacci, A. Rustichini, M. Taboga, \emph{Portfolio selection with monotone mean-variance preferences}, Math. Finance, 19 (2009), 487 -- 521.

\bibitem{Markowitz} H.  Markowitz, \emph{Portfolio selection}, J. Finance, 7 (1952), 77 -- 91.

\bibitem{Mataramvura} S. Mataramvura, B. {\O}ksendal, \emph{Risk minimizing portfolios and HJBI equations for stochastic differential games}, Stochastics, 80 (2008), 317 -- 337.

\bibitem{Oksendal3} B. {\O}ksendal, A. Sulem, \emph{Forward-backward stochastic differential games and stochastic control under model uncertainty}, J. Optim. Theory Appl. 161 22--55.

\bibitem{Oksendal2} B. {\O}ksendal, A. Sulem, \emph{Risk indifference pricing in jump diffusion markets}, Math. Finance, 19 (2009), 619 -- 637. 

\bibitem{Pham2}  H. Pham,  \emph{Continuous-time Stochastic Control and Optimization with Financial Applications}, Stochastic Modelling and Applied Probability, Springer-Verlag, Berlin, 2009.

\bibitem{Shen} Y. Shen, X. Zeng,  \emph{Optimal investment strategy for mean variance insurers with square root factor
process}, Insur. Math. Econ. 62 (2015), 118 -- 137.

\bibitem{Neumann} J. von Neumann , O. Morgenstern,   \emph{Theory of games and economic behavior},  (Princeton: Princeton
university press), 1944.

\bibitem{Yaari} M. Yaari, \emph{The dual theory of choice under risk}, Econometrica 55 (1987), 95--115.

\bibitem{Zari2} T. Zariphopoulou, \emph{A solution approach to valuation with unhegeable risks}, Finance  Stoch. 5 (2001), 61 -- 82.

\bibitem{Zawisza1}  D. Zawisza, \emph{Robust portfolio selection under exponential preferences}, Appl. Math. 37 (2010) (Warsaw), 215 -- 230.

\bibitem{Zawisza2}  D. Zawisza, \emph{Target achieving portfolio under model misspecification: quadratic optimization framework}, Appl. Math. (Warsaw) 39 (2012), 425 -- 443.

\bibitem{Taksar} X. Zeng, M. Taksar  \emph{A stochastic volatility model and optimal portfolio selection}, Quant. Finance
13, 1547 -- 1558.

\bibitem{Zhou} X. Y. Zhou, D. Li, \emph{Continuous-time mean-variance portfolio selection: a stochastic LQ framework},  Appl. Math. Optim. 42 (2000), 19 -- 33.


\end{thebibliography}



\end{document}